\documentclass[iop]{emulateapj}
\usepackage{appendix}
\pdfoutput=1

\begin{document}

\newcommand{\mirlum}{L_{\rm 8}}
\newcommand{\ebmv}{E(B-V)}
\newcommand{\lha}{L(H\alpha)}
\newcommand{\lir}{L_{\rm IR}}
\newcommand{\lbol}{L_{\rm bol}}
\newcommand{\luv}{L_{\rm UV}}
\newcommand{\rs}{{\cal R}}
\newcommand{\ugr}{U_{\rm n}G\rs}
\newcommand{\ks}{K_{\rm s}}
\newcommand{\gmr}{G-\rs}

\title{The MOSDEF Survey: Measurements of Balmer Decrements and the
  Dust Attenuation Curve at Redshifts $z\sim
  1.4-2.6$\,\altaffilmark{*}}

\author{\sc Naveen A. Reddy\altaffilmark{1,5}, Mariska
  Kriek\altaffilmark{2}, Alice E. Shapley\altaffilmark{3}, William
  R. Freeman\altaffilmark{1}, Brian Siana\altaffilmark{1}, Alison
  L. Coil\altaffilmark{4}, Bahram Mobasher\altaffilmark{1}, Sedona
  H. Price\altaffilmark{2}, Ryan L. Sanders\altaffilmark{3}, \& Irene
  Shivaei\altaffilmark{1}}

\altaffiltext{*}{Based on data obtained at the W.M. Keck
Observatory, which is operated as a scientific partnership among the
California Institute of Technology, the University of California, and
NASA, and was made possible by the generous financial support of the
W.M. Keck Foundation.}

\altaffiltext{1}{Department of Physics and Astronomy, University of
  California, Riverside, 900 University Avenue, Riverside, CA 92521,
  USA} 
\altaffiltext{2}{Astronomy Department, University of California, Berkeley,
CA 94720, USA}
\altaffiltext{3}{Department of Physics \& Astronomy, University of 
California, Los Angeles, 430 Portola Plaza, Los Angeles, CA 90095, USA}
\altaffiltext{4}{Center for Astrophysics and Space Sciences, University
of California, San Diego, 9500 Gilman Dr., La Jolla, CA 92093-0424, USA}
\altaffiltext{5}{Alfred P. Sloan Research Fellow}

\slugcomment{DRAFT: \today}
\shorttitle{MOSDEF: The Dust Attenuation Curve at $z\sim 2$}
\shortauthors{N. Reddy, M. Kriek, A. Shapley, et~al.}

\begin{abstract}

We present results on the dust attenuation curve of $z\sim 2$ galaxies
using early observations from the MOSFIRE Deep Evolution Field
(MOSDEF) survey.  Our sample consists of 224 star-forming galaxies
with nebular spectroscopic redshifts in the range $z_{\rm spec} = 1.36
- 2.59$ and high $S/N$ measurements of, or upper limits on, the
H$\alpha$ and H$\beta$ emission lines obtained with the MOSFIRE
spectrograph on the Keck~I telescope.  Using deep multi-wavelength
photometry, we construct composite spectral energy distributions
(SEDs) of galaxies in bins of specific star-formation rate
(SFR/$M^{\ast}$) and Balmer optical depth.  These composites are used
to directly constrain the shape and normalization of the dust
attenuation curve over the full wavelength range from the UV through
near-IR for typical star-forming galaxies at high redshift ($z\ga
1.4$).  Our results imply an attenuation curve that is very similar in
shape and normalization to the SMC extinction curve at wavelengths
$\lambda \ga 2500$\,\AA.  At shorter wavelengths, the shape of the
curve is identical to that of the \citet{calzetti00} starburst
attenuation relation, but with a lower normalization ($R_V$), implying
less attenuation at a fixed wavelength for a given SED shape.  Hence,
the new attenuation curve results in SFRs that are $\approx 20\%$
lower, and stellar masses that are $\Delta \log (M^{\ast}/M_\odot)
\simeq 0.16$\,dex lower, than those obtained with the starburst
attenuation curve.  We find marginal evidence for excess absorption at
$2175$\,\AA.  Moreover, we find that the difference in the
reddening---and the total attenuation---of the ionized gas and stellar
continuum correlates strongly with SFR, such that for dust-corrected
SFRs $\ga 20$\,$M_\odot$\,yr$^{-1}$, assuming a \citet{chabrier03}
IMF, the nebular emission lines suffer an increasing degree of
obscuration relative to the continuum.  A simple model that can
account for these trends is one in which the UV through optical
stellar continuum is dominated by a population of less reddened stars,
while the nebular line and bolometric luminosities become increasingly
dominated by dustier stellar populations for galaxies with large SFRs,
as a result of the increased dust enrichment that accompanies such
galaxies.  Consequently, UV- and SED-based SFRs may underestimate the
total SFR at even modest levels of $\approx 20$\,$M_\odot$\,yr$^{-1}$.

\end{abstract}


\section{INTRODUCTION}
\label{sec:intro}

As a fundamental byproduct of star formation, dust plays an important
role in our understanding of heavy element synthesis in stars, the
dispersal of such elements into the ISM, and our interpretation of the
stellar populations of galaxies given the wavelength-dependent effects
of dust absorption.  The latter aspect is encoded in the shape of the
attenuation curve \citep{draine07}, which conveniently parameterizes
the complexities of the dust properties and spatial distribution of
that dust with respect to the stars in galaxies.  Observations of the
Milky Way and nearby galaxies have yielded detailed information on the
shape of the dust extinction and attenuation curves, respectively, in
the local Universe (e.g., \citealt{prevot84, cardelli89, pei92,
  calzetti94, calzetti00, gordon03, johnson07b, wild11}).  Armed with
such information, one can then deduce the amount of attenuation at any
given wavelength (i.e., the dust obscuration) and recover the
bolometric star-formation rate (SFR), accounting for the starlight
absorbed and re-radiated by dust.

Understanding the shape of this curve takes on added importance for
evaluating the global energetics and stellar populations of galaxies
at high redshift, where direct measures of dust emission from
far-infrared (far-IR) data become observationally inaccessible for
typical star-forming galaxies (e.g., \citealt{reddy06a}).  The
stacking of mid- and far-IR data from the {\em Spitzer Space
  Telescope} and the {\em Herschel Space Observatory} has proved
beneficial for our understanding of the {\em average} dust properties
of typical $L^{\ast}$ star-forming galaxies at $z\sim 2$
\citep{reddy06a, daddi07a, reddy12a}.  IR observations of lensed
star-forming galaxies can lend insight into the dust emission from
individual $L^{\ast}$ (and fainter) galaxies (e.g., \citealt{siana08,
  siana09, sklias14}), but these samples remain small.  These stacking
and lensed galaxy studies lack the precision required to uncover
changes in dust properties and dust/stars geometry from galaxy to
galaxy.  Further, while the UV continuum slope (e.g.,
\citealt{calzetti94, meurer99}) has been shown by numerous studies to
be a reasonable proxy for dust attenuation in star-forming galaxies in
the local and high-redshift universe (e.g., \citealt{calzetti94,
  meurer99, adelberger00, nandra02, reddy06a, daddi07a, reddy10,
  overzier11, reddy12a, buat12}, among many others), and while it can
be measured for individual galaxies to very faint magnitudes at
high redshift \citep{bouwens12, finkelstein12, dunlop12, oesch13}, it
is nonetheless also sensitive to age, metallicity, and star-formation
history (e.g., \citealt{kong04, seibert05, johnson07a, dale09,
  reddy10, boquien12, wilkins13}; see also below).  In addition, the
measurement of the UV slope can be complicated by the presence of the
$2175$\,\AA\, absorption feature \citep{noll09, buat11, kriek13, buat12}.
Despite these issues, the UV slope is by far the most heavily used
dust indicator at high redshift given that it can be measured easily
for individual star-forming galaxies, particularly those selected by
their rest-frame UV colors.

Separately, existing studies of the shape of the attenuation curve at
$z\ga 1.5$ have relied either on spectroscopy over limited wavelength
ranges, indirect inferences based on SED modeling of broad- and
medium-band data and/or SFR comparisons \citep{reddy06a, erb06c,
  noll09, reddy10, buat11, buat12, kriek13}, or have been investigated
for galaxies over a limited range of spectral shape (e.g., galaxies
with weak Balmer and 4000\,\AA\, breaks) and UV luminosities with an
explicit assumption for the shape of the intrinsic spectrum
\citep{scoville15}.  What have been needed for a more robust study of
the attenuation curve are indicators of dust and SFR at high redshift
that: (a) can be measured for individual $L^{\ast}$ galaxies at high
redshift; (b) are less attenuated than the UV luminosity; (c) are less
sensitive to age and star-formation history; and (c) probe star
formation on the shortest timescales.  When coupled with samples of
galaxies with near-IR spectra and broad- and intermediate-band
photometry spanning the UV through near-IR wavelengths, these SFR/dust
indicators can be used to address the shape of the attenuation curve
of high-redshift galaxies.

The HI recombination line luminosities and their ratios have been
considered the ``gold standard'' for measuring SFRs and dust
attenuation in local galaxies \citep{kennicutt12}.  In particular, the
Balmer (and higher series) transitions in star-forming galaxies arise
from the HII regions generated by the most massive O stars ($\ga
8$\,$M_\odot$) and therefore trace the SFR averaged over timescales of
just a few Myr, making these tracers insensitive to the star-formation
history on longer timescales.  In nearby galaxies, these lines are
easily observable from the ground to faint limits, and they are less
affected by attenuation than UV emission.  The higher series lines,
including Pa$\alpha$ ($\lambda = 1.875$\,$\mu$m) and Br$\gamma$
($\lambda = 2.166$\,$\mu$m), are particularly useful indicators as
their relatively long wavelengths imply negligible dust absorption.
However, these features are weak relative to the Balmer lines and, at
$z\ga 1.5$, shift into the mid-IR where current instrumentation has been
insufficient to detect the lines for all but the most luminous and
strongly gravitationally lensed galaxies \citep{papovich09}.

Until recently, obtaining measurements of nebular recombination (other
than Ly$\alpha$) has been prohibitive for faint galaxies at redshifts
$z\ga 0.5$ as these lines are shifted into the near-infrared (near-IR)
where the terrestrial background is substantially higher than in the
optical, with a color of $(B-\ks)\simeq 7$\,mag.  Deep ground-based
narrowband imaging \citep{geach08, lee12, ly12, sobral12, momcheva13},
and space-based grism spectroscopy afforded by the {\em Hubble Space
  Telescope} ({\em HST}) WFC3 camera \citep{atek10, brammer12,
  dominguez13, price14} have begun to advance the study of nebular
line emission (particularly H$\alpha$ and H$\beta$) in
intermediate-redshift galaxies ($0.5\la z\la 2.5$).  Nonetheless, such
studies are generally limited to detecting high equivalent width
lines, do not allow for the separation of lines close in wavelength
(e.g., H$\alpha$ and [NII]), or have a low resolution ($R<130$) with a
limited wavelength range that hinders the determination of physically
important line ratios and widths.  Moderate-resolution near-IR
spectroscopy exists for both UV-selected star-forming galaxies
\citep{erb06b, mannucci09} and near-IR-selected samples of massive
($M\ge 10^{11}$\,$M_\odot$) galaxies \citep{kriek08}, yet these
samples are quite small and typically do not cover both the $H\alpha$
and H$\beta$ lines.  In cases where both lines were targeted (e.g.,
\citealt{kriek08}), the H$\beta$ line is commonly undetected given the
depths of previous spectroscopic campaigns.  The advances enabled with
the new generation of multi-object near-IR spectrographs on
8-10\,m-class telescopes have begun to remedy these past deficiencies
in high-redshift near-IR spectroscopy (e.g., \citealt{kashino13,
  steidel14}).

To this end, we have begun the MOSFIRE Deep Evolution Field (MOSDEF)
survey \citep{kriek14}, conducted with the recently-commissioned
multi-object near-IR spectrograph MOSFIRE \citep{mclean12} on the
Keck~I telescope.  MOSDEF aims to characterize the SFRs, dust
attenuation, chemical content, ISM physical conditions, dynamics,
stellar populations, and black hole accretion activity of a large
sample of $\sim 1500$ {\em H}-selected galaxies and AGN in the
redshift range $z=1.4-3.8$, corresponding to the peak of star
formation (see review of \citealt{madau14}) and active galactic
nucleus (AGN) activity (e.g., \citealt{silverman08}).  In this study,
we use a sample of 224 galaxies with measurements of the H$\alpha$ and
H$\beta$ emission lines from the first two years of the MOSDEF survey
to provide a direct measurement of the shape of the attenuation curve
over the full wavelength range from the UV through near-IR for typical
star-forming galaxies at $z\ga 1.4$.  Our analysis marks a significant
improvement over existing studies because: (a) we consider galaxies
with a large range in spectral shapes (from very blue and dust-free
star-forming galaxies to very dusty and highly star-forming ones); (b)
we use an indicator of dust, namely the Balmer decrement, which is
sensitive to the reddening of the ionized gas and thus is physically
distinct from the shape of the SED (and thus less sensitive to the
details of the star-formation history and age of the stellar
population), and which can be measured for individual galaxies in our
sample; (c) our method of computing the attenuation curve is
model-independent in that we make no assumption of the shape of the
instrinic SED---rather, we use the data themselves to establish the
how galaxy spectral shapes vary with increased dust attenuation; and
(d) the extensive multi-wavelength photometry affords us the ability
to measure the shape of the attenuation curve at long wavelengths and,
thus, the normalization of the curve.

In Section~\ref{sec:sample}, we introduce the MOSDEF survey and the
basic measurements including line fluxes, stellar population
parameters, and SFRs.  Section~\ref{sec:effectofdust} presents a
discussion of the effects of dust on the stellar continuum and the
derivation of the attenuation curve.  The implications for the
attenuation curve on our inferences of the stellar populations and
SFRs, as well as the total attenuation of the stellar continuum and
ionized gas, in distant galaxies are discussed in
Section~\ref{sec:discussion}.  Throughout we assume a
\citet{chabrier03} initial mass function (IMF) and a cosmology with
$H_{0}=70$\,km\,s$^{-1}$\,Mpc$^{-1}$, $\Omega_{\Lambda}=0.7$, and
$\Omega_{\rm m}=0.3$.  Line wavelengths are in vacuum.  All magnitudes
are expressed in the AB system \citep{oke83}.

\section{Data and Basic Measurements}
\label{sec:sample}

\subsection{MOSDEF Survey}

The MOSDEF survey is being conducted using the MOSFIRE spectrograph
\citep{mclean12, steidel14} on the 10\,m Keck I telescope.  This
recently commissioned instrument enables the simultaneous near-IR
spectroscopy of up to $\approx 46$ (with typically 30) galaxies across
a $6\arcmin \times 3\arcmin$ field of view.  MOSFIRE provides a
wavelength coverage from $0.97$ to $2.40$\,$\mu$m with a spectral
resolution of $R\approx 3300$, $3650$, and $3600$ in the {\em J}, {\em
  H}, and {\em K} bands, respectively.  Details of the survey,
observations, and data reduction are provided in \citet{kriek14}, and
below we summarize the salient aspects relevant for this analysis,
which uses the current data from the first two observing seasons of
the survey.

\subsection{Galaxy Selection}

Catalogs including ground- and space-based photometry from $0.3$ to
$8.0$\,$\mu$m, provided to us by the 3D-HST grism survey team (PI: van
Dokkum; \citealt{brammer12, skelton14}), were used for target
selection \citep{kriek14}.  These catalogs are based on sources
detected in a weighted combination of the {\em HST} F125W, F140W, and
F160W images in the five fields of the Cosmic Assembly Near-Infrared
Deep Extragalactic Legacy Survey (CANDELS; \citealt{koekemoer11,
  grogin11}): COSMOS, EGS, GOODS-North, GOODS-South, and UDS.  The
{\em targeting} catalogs include photometric redshifts and, where
available, {\em HST} grism redshifts \citep{brammer12, skelton14} and
external spectroscopic redshifts.  We restricted the targeting
catalogs to include only those galaxies with photometric and/or grism
or external spectroscopic redshifts in the three ranges $z=1.37-1.70$,
$2.09-2.61$, and $2.95-3.80$, where the various strong nebular
emission lines of interest, including [OII]$\lambda\lambda 3727,
3730$\,\AA, [OIII]$\lambda\lambda 4364,5008$\,\AA, H$\beta$,
H$\alpha$, and [NII]$\lambda\lambda 6550, 6585$\,\AA, fall in the {\em
  Y}, {\em J}, {\em H}, and {\em K} atmospheric windows.  Targets are
selected to fixed {\em H-}band magnitude limits of $m_{\rm
  F160W}=24.0$, $24.5$, and $25.0$, for the $z=1.37-1.70$,
$2.09-2.61$, and $2.95-3.80$ redshift ranges, respectively, to ensure
a roughly consistent stellar mass limit of $\approx 10^{9}$\,$M_\odot$
across the three redshift ranges.  For brevity, we refer to these as
the $z\sim 1.5$, $2.3$, and $3.3$ samples.  In this analysis, we
consider only galaxies from the two lower redshift ranges, where both
the H$\alpha$ and H$\beta$ lines are covered.

\subsection{Observations and Efficiency}

\begin{figure*}[t]
\plotone{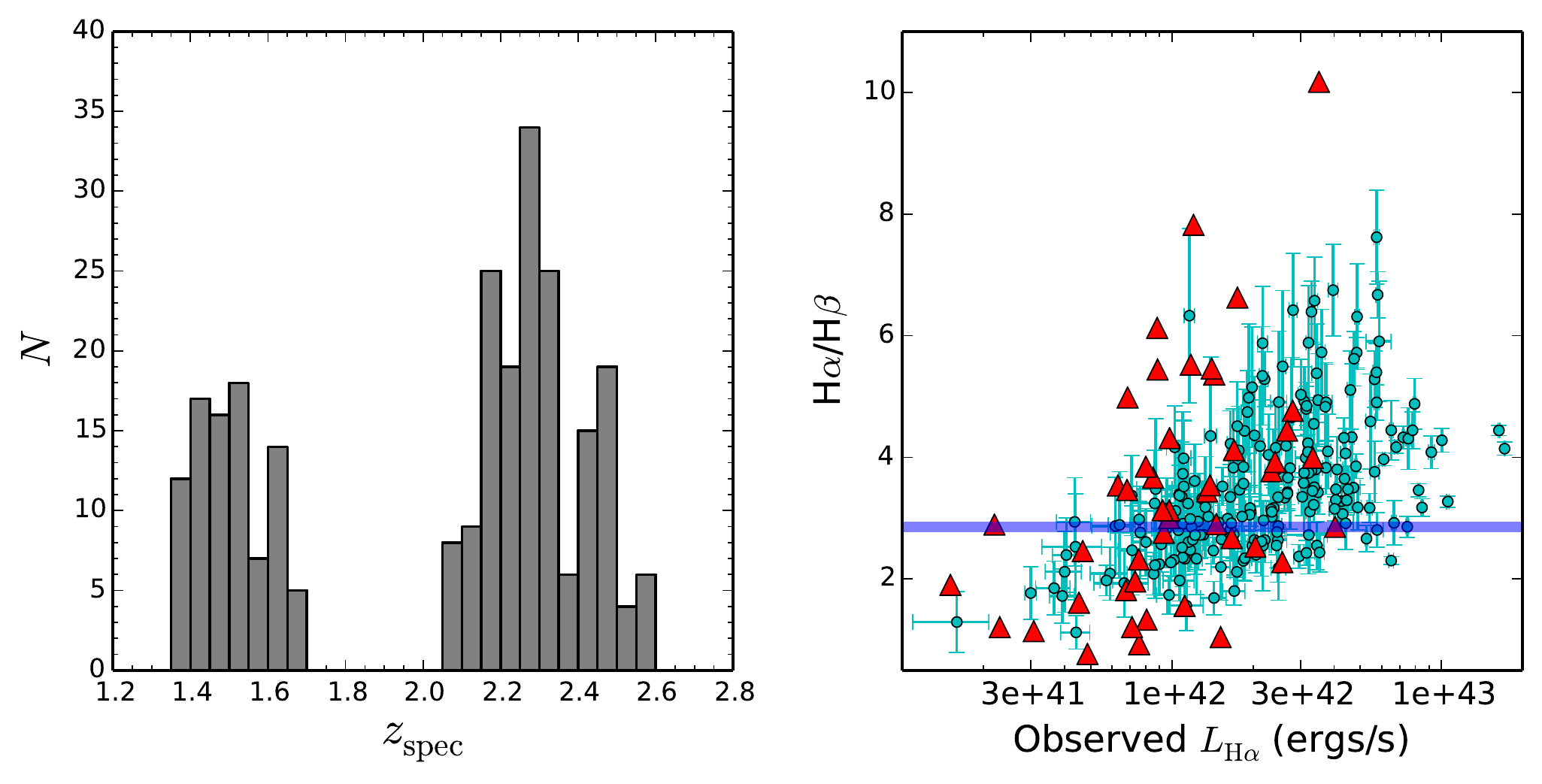}
\caption{{\em Left:} Spectroscopic redshift distribution of 259
  galaxies in the two lower redshift ranges ($z=1.37-1.70$ and
  $2.09-2.61$) targeted with MOSDEF during the first two observing
  seasons.  {\em Right:} Ratio of H$\alpha$ to H$\beta$ emission flux
  versus observed H$\alpha$ luminosity, for our sample of 259 galaxies
  with spectroscopic redshifts $1.35< z_{\rm spec} < 2.59$ and
  $S/N_{\rm H\alpha}\ge 3$.  The emission line fluxes are corrected
  for underlying Balmer absorption.  The 46 galaxies with 3\,$\sigma$
  upper limits in H$\beta$ flux are denoted by the triangles.  The
  horizontal blue line indicates the intrinsic ratio
  (H$\alpha$/H$\beta$ = 2.86) for typical ISM conditions
  \citep{osterbrock89}.  Of the 213 galaxies with significant
  detections of both H$\alpha$ and H$\beta$, 58 and 85 are consistent
  with having non-zero nebular attenuation at the $3$\,$\sigma$ and
  $2$\,$\sigma$ levels, respectively.}
\label{fig:bd_vs_lha}
\end{figure*}

Targets were prioritized for spectroscopic observations based on two
primary criteria.  First, bright galaxies were up-weighted relative to
fainter galaxies to ensure a statistical representation of rarer,
massive galaxies.  Second, within a given magnitude range, objects
were prioritized according to whether they had external spectroscopic
redshifts, grism redshifts, photometric redshifts toward the center of
the redshift ranges specified above, and photometric redshifts at the
edges of the redshift ranges, in that order.  The sample
characteristics, success rate, and comparison of the spectroscopic
sample to the parent sample are presented in \citet{kriek14}.  Masks
were observed using an ABA'B' dither pattern with $0\farcs 7$ width
slits for $2$\,hrs each in the {\em J}, {\em H}, and {\em K} filters
for the $z\sim 2.3$ sample; and $1$\,hr each in the {\em Y}, {\em J},
and {\em H} filters for the $z\sim 1.5$ sample.  Our observations were
taken with a typical seeing of $0\farcs 4 - 1\farcs 0$.  Data
reduction and spectral extraction were accomplished using a custom IDL
pipeline and interactive software \citep{kriek14}.  The prioritization
and exposure time strategy have proved highly successful, with a $\sim
80\%$ spectroscopic confirmation rate on average per mask, and a $\ga
70\%$ success rate for objects at the faint limit of our $z\sim 2.3$
sample.  Figure~\ref{fig:bd_vs_lha} shows the spectroscopic redshift
distribution of 259 galaxies (excluding X-ray and IR-selected AGN;
\citealt{coil15}) in the two lower redshift ranges targeted with
MOSDEF thus far.

\subsection{Slit Loss Corrections}

Slit loss corrections have a particular bearing on our analysis, as
they are required to ensure accurate flux ratios between lines taken
through different photometric filters and observed under different
weather conditions.  Response corrections of the spectra were
performed using telluric standard star observations, and an initial
flux calibration is implemented by forcing the spectra of a bright
``slit star'' placed on each of the science masks to match the star's
broadband photometry.  However, the typical seeing of our observations
is such that most of the targeted galaxies are spatially resolved and
thus cannot be treated as point sources when correcting for the loss
of flux outside the slit apertures.  Thus, we modeled the F160W light
profiles of each of our galaxies using a two-dimensional rotated
elliptical Gaussian which, when convolved with the seeing disk and
integrated through the slit, yields an estimate of the fraction of
flux lost outside the slit, relative to the fraction lost for the slit
star.  The efficacy of this procedure is evaluated by comparing the
spectroscopic magnitudes of galaxies detected in the continuum with
their broadband magnitudes.  Our slit loss correction procedure yields
an absolute spectral flux calibration that is typically within $18\%$
of that predicted from the photometry, with a dispersion of roughly
the same order.

\subsection{Line Flux Measurements}

Line measurements were performed by first estimating an initial
redshift based on the highest signal-to-noise ($S/N$) line for each
galaxy, and then fitting Gaussian functions at the observed
wavelengths of all other lines.  We allowed for a linear fit to the
continuum underlying each line, and the [OII]$\lambda 3727, 3730$ and
H$\alpha$+[NII]$\lambda 6550, 6585$ lines were fit using double and
triple Gaussian functions, respectively.  To estimate line flux
errors, the spectra of each galaxy were perturbed 1000 times according
to their error spectra and all the lines were remeasured from these
realizations.

The H$\alpha$ and H$\beta$ line fluxes were corrected for underlying
Balmer absorption arising primarily from the atmospheres of A stars.
As the continuum is generally undetected in the MOSDEF spectra, we
estimated the Balmer absorption from the stellar population model fits
to the galaxies, as discussed in the next section.  For each model
fit, we measured the absorbed flux relative to the continuum.
Generally, the corrections reach peak rest-frame equivalent widths of
$\la 3$\,\AA\, and $\la 4$\,\AA\, for H$\alpha$ and H$\beta$,
respectively, at an age of $500$\,Myr.  For a constant star-formation
history, the equivalent widths equilibrate to $\approx 2-3$\,\AA\, for
older ages.  The typical corrections to the H$\alpha$ and H$\beta$
emission line fluxes are $\la 3\%$ and $\la 10\%$, respectively.  The
corrected H$\alpha$/H$\beta$ ratios for our sample are shown in
Figure~\ref{fig:bd_vs_lha}.  The random uncertainties in the Balmer
absorption line fluxes were estimated by calculating the dispersion in
the absorbed fluxes for model stellar population fits to the
photometry perturbed many times according to the photometric errors.
The uncertainties in the absorbed fluxes are $\simeq 2\%$ and much
smaller than the typical $\approx 15\%$ uncertainties in the emission
line flux measurements, and therefore we chose not to include the
additional random error of the absorbed fluxes in our analysis.

It is useful to assess whether the 46 galaxies with $3$\,$\sigma$
upper limits in H$\beta$ have distinctive properties relative to those
of the 213 H$\beta$-detected galaxies.  The 46 H$\beta$-undetected
galaxies have lower limits in H$\alpha$/H$\beta$ that span the full
range observed for the H$\beta$-detected galaxies, and the broadband
SEDs for the former are otherwise unremarkable compared to the latter.
Moreover, the frequency with which the H$\beta$ line is significantly
affected by strong skylines is much higher (37 out of 46, or $80\%$)
for the H$\beta$-undetected galaxies than for the H$\beta$-detected
galaxies (60 out of 213, or $28\%$).  Thus, the H$\beta$
non-detections can at least partly be attributed to the fact that the
galaxies lie at redshifts where H$\beta$ happened to fall on or close
to a strong skyline.

To further investigate any systematic differences between these 46
galaxies and the rest of the sample, we constructed weighted average
spectra of the H$\beta$-detected and -undetected galaxies and measured
the H$\alpha$ and H$\beta$ fluxes in these combined spectra.  The
weighting of individual spectra when computing the average is simply
$1/\sigma(\lambda)^2$, where $\sigma(\lambda)$ if the value of the
error spectrum at wavelength $\lambda$.  The H$\beta$-detected and
-undetected galaxies have $\langle {\rm H}\alpha/{\rm H}\beta\rangle =
4.1$ and $4.3$, respectively, after correcting for the mean Balmer
absorption in these galaxies.  This difference is smaller than the
dispersion in the mean Balmer decrements obtained when we bootstrap
resample the galaxies contributing to the mean stacks.  The galaxies
with undetected H$\beta$ have an average Balmer decrement that is
similar to that of the H$\beta$-detected galaxies, implying that they
suffer the same degree of obscuration on average.  Only those galaxies
with $>3$\,$\sigma$ detection of H$\beta$ (and H$\alpha$) are
considered in the derivation of the attenuation curve, but we
specifically include the sample of H$\beta$-undetected galaxies
throughout the paper as necessary.

\subsection{Stellar Population Modeling}
\label{sec:sedmod}

To aid our analysis, we modeled the stellar populations of galaxies in
our sample to derive SED-based SFRs, color excesses, ages, and stellar
masses.  The basic procedure was to model the multi-wavelength photometry of
the galaxies using a suite of \citet{bruzual03} stellar population
models for different star-formation histories, ages, and reddening;
the latter assumes the \citet{calzetti00} attenuation curve, but we
consider other attenuation curves below.  The photometry is
corrected, if necessary, for the contribution from the strongest
emission lines measured in our survey, including H$\alpha$, H$\beta$,
[OIII]$\lambda\lambda 4960, 5008$, and [OII]$\lambda\lambda 3727,
3730$.  We modeled the stellar populations using constant, declining,
and exponentially rising star-formation histories with e-folding times
of $\tau=100-5000$\,Myr, and generally adopted the exponentially
rising histories as the default for reporting results from the
fitting.  In the case of 14 galaxies, exponentially declining models
were adopted as they yielded better fits to the rest-UV photometry.
Incorporating the parameters obtained for a constant star-formation
history does not alter our analysis or conclusions.  In the modeling,
we allowed the age to vary from $50$\,Myr (approximately the dynamical
timescale; e.g., \citealt{reddy12b}) to the age of the Universe at the
redshift of each galaxy.  Reddening in the range $0.0\le E(B-V) \le
0.6$ was considered.  We adopted the parameters corresponding to the
model that gave a minimum $\chi^2$ relative to the 
photometry.  Errors in the best-fit parameters were derived by
perturbing the photometry assuming a Gaussian distribution
of full-width-half-max equal to the photometric uncertainties.  Each
of these realizations of the photometry were then refit to find the
best-fit model, and the errors in best-fit parameters were taken to be
the values that encompassed $68\%$ of the simulations.

Of the 259 galaxies in our sample, 35 (31 H$\beta$-detected and 4
H$\beta$-undetected) galaxies had noisy photometry relative to the
best-fit model and were excluded from subsquent analysis.  Thus, the
final sample consists of 224 galaxies, with 182 H$\beta$-detected
galaxies and 42 H$\beta$-undetected galaxies.

\subsection{Balmer Optical Depth}

As in \citet{calzetti94}, we define the {\em Balmer optical depth}
\begin{equation}
\tau_{\rm b}\equiv \ln\left(\frac{{\rm H}\alpha/{\rm H}\beta}{2.86}\right).
\end{equation}
Here, $\tau_{\rm b}$ is the {\em difference} in optical depths
measured at the wavelengths of H$\beta$ and H$\alpha$, with the usual
definition: $\tau_{\lambda} = -\ln (I_\lambda/I^0_\lambda)$ where
$I_\lambda$ and $I^0_\lambda$ are the observed and intrinsic
intensities, respectively.  Thus, $\tau_{\rm b}$ is directly related
to the color-excess, $E(B-V)$.  Objects with measured
H$\alpha$/H$\beta < 2.86$, the theoretical minimum value in the
absence of dust for Case B recombination and $T=10000$\,K
\citep{osterbrock89}, are assigned $\tau_{\rm b}=0$.\footnote{The
  intrinsic ratio H$\alpha$/H$\beta$ is insensitive to the electron
  density, varying by less than $2\%$ from $n_e = 10^2$ to
  $10^6$\,cm$^{-3}$.}

\subsection{SFR Calculations}
\label{sec:sfrcalc}

The default measurements we use for the SFRs come not from the SED
fitting, but from the H$\alpha$ flux measurements.  The latter are
more immune to the degeneracies inherent in SED modeling, less
affected by extinction than the rest-UV, and trace the
``instantaneous'' SFR.  We used the \citet{kennicutt98} relation to
convert H$\alpha$ luminosities to SFRs, assuming a \citet{chabrier03}
IMF.  These observed SFRs were then corrected for attenuation by
assuming an attenuation curve ($k(\lambda)$) and nebular color excess
($E(B-V))_{\rm gas}$).  In most previous investigations at {\em high
  redshift}, the attenuation curve assumed for the nebular line
regions is the \citet{calzetti00} curve, with color excesses based on
those found for the stellar continuum (also assuming a
\citealt{calzetti00} curve).  In our analysis, we calculated
SFR(H$\alpha$) assuming the \citet{cardelli89} Galactic extinction
curve, as was the original intent of \citet{calzetti94}, but we
explore the effect of assuming other attenuation curves in deriving
the color excess, as well as the implications for dust corrections and
bolometric SFRs, in Sections~\ref{sec:gasvsstars} and
\ref{sec:sfrimplications} below.  Throughout the paper, and unless
noted otherwise, the specific SFRs (sSFRs) are computed assuming the
dust-corrected SFRs(H$\alpha$) and the SED-determined stellar masses.
This approach is more advantageous than combining SED-determined SFRs
with $M^{\ast}$ as these quantities are highly correlated with each
other (i.e., both SFR(SED) and $M^{\ast}$ are constrained by the
normalization of the best-fit SED model with the photometry).

\section{The Effect of Dust on the Stellar Continuum Emission}
\label{sec:effectofdust}

As noted above, the distribution of H$\alpha$/H$\beta$ ratios implies
the presence of dust in the majority of the objects in our sample.
Before considering the level of obscuration required to reproduce the
observed Balmer decrements---a step that requires the assumption of an
attenuation curve---it is instructive to examine how the galaxy
spectral shapes vary with H$\alpha$/H$\beta$ ratio.  We focus
specifically on the sensitivity of the UV spectral slope to dust
obscuration, given the large absorption cross-section of dust to UV
photons, and the prevalent use of the UV slope as a dust indicator for
typical star-forming galaxies at high redshifts.  Formally, the UV
slope $\beta$ is defined such that $f_\lambda \propto \lambda^\beta$.
Note that $\beta$ is the {\em observed} UV continuum slope,
irrespective of the underlying stellar population.  Thus, a galaxy
with an older stellar population and a significant contribution of UV
flux from stars of spectral type A and later, can exhibit the same
spectral slope as a galaxy whose UV continuum is dominated by OB stars
and contains significant amounts of dust.

\subsection{Calculation of the UV Slope}

In practice, $\beta$ can be measured directly from a high
signal-to-noise spectrum using a number of continuum windows spanning
the range $1250-2600$\,\AA, where the windows are designed to exclude
regions of strong interstellar/stellar absorption and the
$2175$\,\AA\, dust absorption feature \citep{calzetti94}.  For most
galaxies lacking rest-UV spectroscopic data, as in our case, either
broad- and intermediate-band photometry, or the stellar population
model that best fits such photometry, can be used to estimate $\beta$.
We calculated a ``photometric'' UV slope ($\beta_{\rm phot}$) by
fitting directly the photometry (taking into account the
photometric errors), which includes for all objects in our sample
anywhere from 4 to 19 photometric points that lie in the (rest-frame)
range $\lambda = 1250-2600$\,\AA.  The measurement uncertainty in the
rest-UV photometry results in best-fit UV slopes with a typical error
of $\Delta\beta_{\rm phot}\simeq 0.2$

For comparison, we also computed an ``SED'' UV slope ($\beta_{\rm
  SED}$) by fitting the flux of the best-fit stellar population model
as a function of wavelength, using only those flux points that lie
within the 10 continuum windows given in \citet{calzetti94}.  
The typical formal uncertainty in $\beta_{\rm SED}$, when using the 10
aforementioned windows, is $\Delta\beta_{\rm SED} \simeq 0.1$ (not
including the systematic error associated with the assumed stellar
population model).  The two measures of UV slope, $\beta_{\rm phot}$
and $\beta_{\rm SED}$, are highly correlated because the stellar
population model is fit to the same photometry used to calculate
$\beta_{\rm phot}$.

\subsection{Relation between UV slope and Balmer Line Opacity}

Our expectation is that if the UV slope is sensitive to dust
obscuration, which a number of studies have shown to be the case for
different subsets of high-redshift galaxies (e.g., \citealt{nandra02,
  reddy04, reddy06a,daddi07a, pannella09, buat09, magdis10,
  reddy12a}), then we should observe a correlation between $\beta$ and
$\tau_{\rm b}$.  The scatter between these two quantities may hint at
other factors that influence $\beta$, including star-formation
history, metallicity of the underlying stellar population, and
differences in the geometrical distribution of stars and dust in
galaxies.

\begin{figure*}[t]
\epsscale{1.15}
\plotone{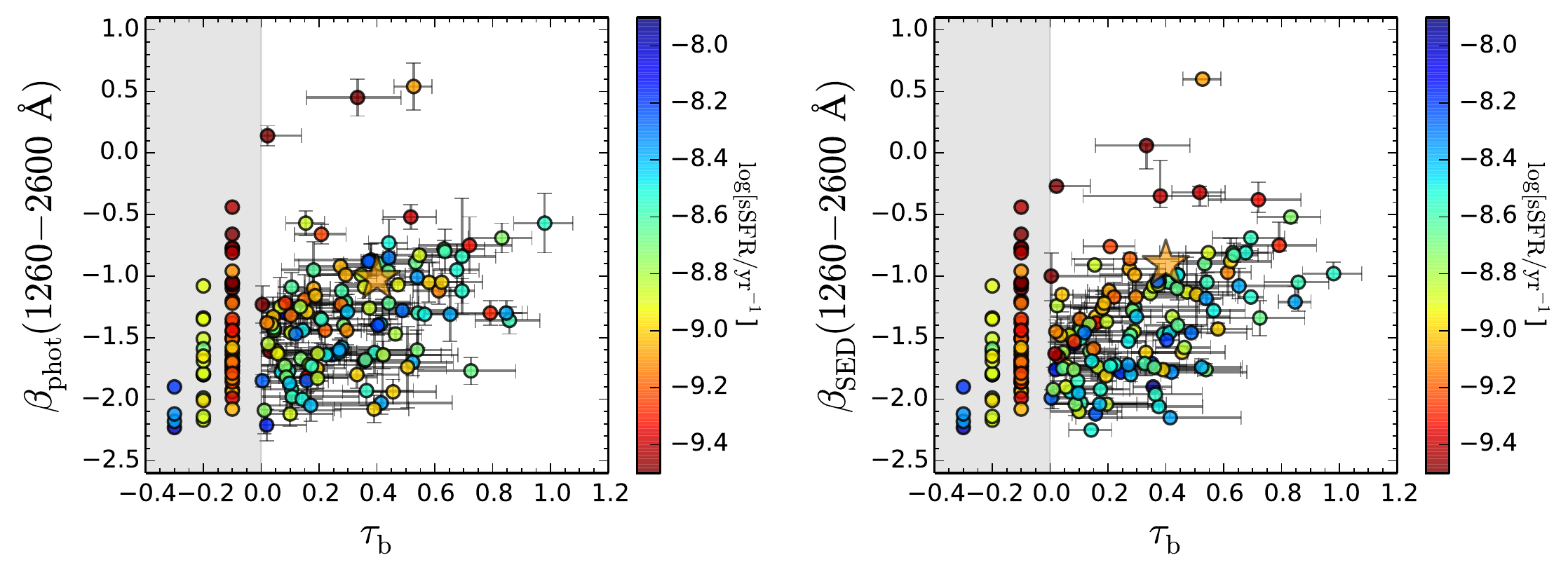}
\caption{Photometric ({\em left}; $\beta_{\rm phot}$) and SED-based
  ({\em right}; $\beta_{\rm SED}$) UV slopes vs. Balmer optical depth
  ($\tau_{\rm b}$), differentiated by sSFR.  Errors in $\tau_{\rm b}$
  include both measurement uncertainties in the H$\alpha$ and H$\beta$
  fluxes, and the uncertainty in slitloss corrections.  Errors in
  $\beta_{\rm phot}$ are derived directly from the linear regression
  between $\log f_\lambda$ and $\lambda$.  Errors in $\beta_{\rm SED}$
  are estimated from Monte Carlo simulations of the SED fitting (see
  text).  The sSFRs assume dust-corrected H$\alpha$-based SFRs and the
  stellar masses returned from the SED fitting.  The large stars
  indicate the average positions of the H$\beta$-undetected galaxies.
  Galaxies with $\tau_{\rm b}=0$ and different ranges of sSFR are
  arbitrarily offset to the left side of each panel for clarity.}
\label{fig:beta_vs_taub}
\end{figure*}

Figure~\ref{fig:beta_vs_taub}a shows the distribution of $\beta_{\rm
  phot}$ with $\tau_{\rm b}$ for the 182 objects in our sample with
$S/N_{\rm H\beta} \ge 3$.  The average H$\alpha$/H$\beta$ ratio for
the 42 H$\beta$-undetected galaxies, as calculated from the composite
spectrum (see above), is $4.1\pm0.1$, corresponding to $\tau_{\rm b} =
0.37\pm 0.02$.  The average value of the {\em individual} $\beta_{\rm
  phot}$ measurements for these 42 galaxies is $\beta_{\rm
  phot}=-1.00$.  These average values for the H$\beta$-undetected
galaxies are indicated by the star in Figure~\ref{fig:beta_vs_taub}a.
Generally, there is a large dispersion in $\beta_{\rm phot}$ at a
given Balmer optical depth.  In the limit where $\tau_{\rm
  b}\rightarrow 0$, $\beta$ varies from the bluest ($\beta_{\rm
  phot}\simeq -2.3$) to reddest ($\beta_{\rm phot}\simeq -0.5$) values
observed.  As noted above, $\beta$ can be sensitive to factors other
than the dust content, possibly accounting for some of the scatter in
UV slope at a given Balmer optical depth.  These factors are discussed
in detail below.

\subsection{Effect of Star-formation History on the UV Slope}

\subsubsection{Variations with Specific Star-Formation Rate}
\label{sec:variations_ssfr}

Galaxies with significant older stellar populations can exhibit UV
slopes that are redder than those predicted from dust attenuation
alone.  Specifically, A stars from previous generations of star
formation can contribute significantly to the near-UV flux if the
current SFR---and hence the contribution from O and B stars---is low,
as might be the case for galaxies with low specific SFRs (sSFRs).  The
two consequences of this effect are to increase the scatter in $\beta$
at a fixed $\tau_{\rm b}$, and to accentuate the trend in $\beta$
vs. $\tau_{\rm b}$ if dustier galaxies are on average older and more
massive.  For the moment, we consider the effect of star-formation
history (i.e., effect of older stellar populations) on $\beta$ by
color-coding the points in Figure~\ref{fig:beta_vs_taub} by sSFR.  It
is apparent that there is a ``sequence'' in the location of galaxies
in the $\beta_{\rm phot}$-$\tau_{\rm b}$ plane with sSFR, such that
the locus of points shifts towards redder $\beta_{\rm phot}$ at a
fixed $\tau_{\rm b}$ with {\em decreasing} sSFR.  This is the expected
behavior if late-type stars increasingly affect the near-UV flux with
decreasing sSFR.  The shifting of the locus is also (unsurprisingly)
apparent if we differentiate the points again by sSFR, but use
$\beta_{\rm SED}$ instead of $\beta_{\rm phot}$
(Figure~\ref{fig:beta_vs_taub}).

To demonstrate quantitatively the shifting of the $\beta$
vs. $\tau_{\rm b}$ relation with sSFR, we executed two tests.  In the
first, we used K-S statistical tests to determine the probability that
two distributions of $\beta$ at two given sSFRs (and at a fixed
$\tau_{\rm b}$) are drawn from the same parent distribution.  In
practice, we gridded the data in three bins of $\tau_{\rm b}$
($\tau_{\rm b}$=0, 0.0-0.3, and 0.3-1.0), and two bins of $\log({\rm
  sSFR/yr^{-1}})$, where $-9.60 \le \log({\rm sSFR/yr^{-1}}) \le
-8.84$ and $-8.84 \le \log({\rm sSFR/yr^{-1}}) \le -8.00$.  The sSFR
bin boundaries were chosen to encompass most of the objects in our
sample, with an equal number of objects per sSFR bin.  We then
performed two-sided K-S tests on the two distributions of $\beta$ (for
the two sSFR bins) in each bin of $\tau_{\rm b}$.  The results are
summarized in Table~\ref{tab:kstwo}, and they suggest that the $\beta$
distributions between the two sSFR bins are drawn from different
parent populations.

\begin{deluxetable}{cccc}
\tabletypesize{\footnotesize}
\tablewidth{0pc}
\tablecaption{K-S Test Results\tablenotemark{a}}
\tablehead{
\colhead{} &
\colhead{$\tau_{\rm b}=0$} &
\colhead{$\tau_{\rm b}=0.0-0.3$} &
\colhead{$\tau_{\rm b}=0.3-1.0$}}
\startdata
$\beta_{\rm phot}$ & 0.06 & 0.008 & 0.09 \\
$\beta_{\rm SED}$ & 0.05 & 0.004 & 0.05
\enddata
\tablenotetext{a}{Probability that the distributions of $\beta$ in the
two bins of $\log({\rm sSFR/yr^{-1}})$ ($-9.60 \le \log({\rm sSFR/yr^{-1}}) \le
-8.84$ and $-8.84 \le \log({\rm sSFR/yr^{-1}}) \le -8.00$)
in the
given bins of $\tau_{\rm b}$ are drawn from the same parent population.}
\label{tab:kstwo}
\end{deluxetable}

\begin{figure}[t]
\plotone{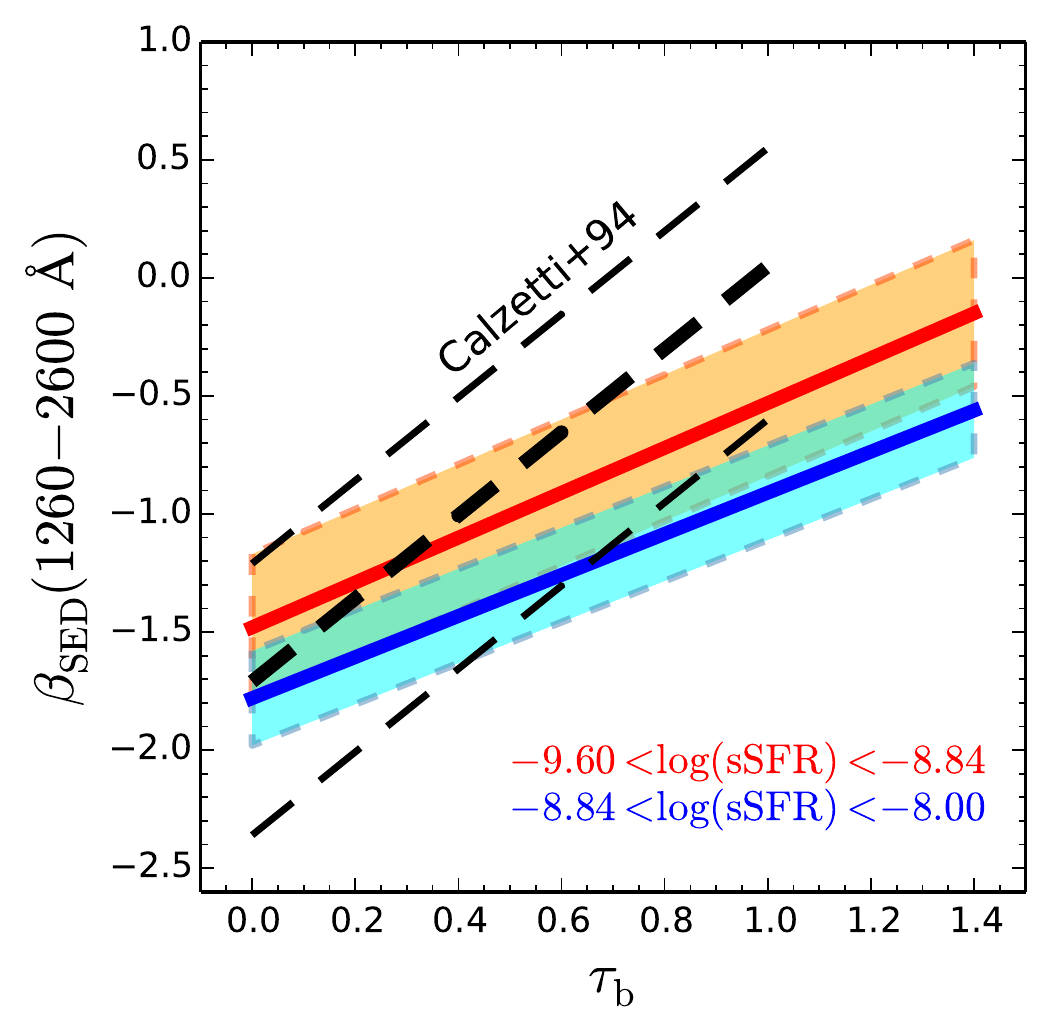}
\caption{Linear regressions between $\beta_{\rm SED}$ and $\tau_{\rm
    b}$.  The solid lines and shaded regions indicate the best-fit
  linear functions and estimated $\pm \sigma_{\rm int}$ intrinsic
  scatter for the different bins of sSFR, respectively.  Individual
  data points have been omitted for clarity. Long-dashed lines
  indicate the best-fit linear function and upper and lower envelopes
  from \citet{calzetti94} for a sample of 39 local starburst and blue
  compact galaxies \citep{kinney93}.}
\label{fig:linfit}
\end{figure}

In the second test, we divided the data into the same two bins of sSFR
defined above, and we computed the Spearman correlation coefficient
and the significance with which the null hypothesis of no correlation
between $\beta$ and $\tau_{\rm b}$ can be ruled out.
Table~\ref{tab:linfit} summarizes the results for $\beta_{\rm SED}$
vs. $\tau_{\rm b}$ and $\beta_{\rm phot}$ vs. $\tau_{\rm b}$.
Significant correlations between $\beta$ and $\tau_{\rm b}$ are found
in most cases, with the slopes and intercepts from linear regressions
indicated in Table~\ref{tab:linfit}.  We also estimated the intrinsic
dispersion in the data ($\sigma_{\rm int}$) by assuming that the total
dispersion in $\beta$ vs. $\tau_{\rm b}$ is the quadrature sum of
$\sigma_{\rm int}$ and the measurement error, $\sigma_{\rm meas}$.
Clearly, $\sigma_{\rm int}$ will depend on the adopted bin widths, but
our goal was to determine the degree to which the $\beta-\tau_{\rm b}$
correlations in given sSFR bins overlap when considering the intrinsic
scatter in the data.  The fits obtained for $\beta_{\rm SED}$
vs. $\tau_{\rm b}$ are shown in Figure~\ref{fig:linfit}.  From these
statistical tests, it is clear that the scatter seen in
Figure~\ref{fig:beta_vs_taub} is to a large extent due to variation in
the $\beta$ vs. $\tau_{\rm b}$ relations for galaxies with different
sSFRs.  Specifically, the relation for the lowest sSFR galaxies is
shifted to redder $\beta$ by $\approx 2\times \sigma_{\rm int}$
relative to the relation for the highest sSFR galaxies.

\begin{deluxetable*}{ccccccccc}
\tabletypesize{\footnotesize}
\tablewidth{0pc}
\tablecaption{Spearman Correlation Test and Linear Regression Results for $\beta$ vs. $\tau_{\rm b}$}
\tablehead{
\colhead{$\beta$} &
\colhead{$\log[{\rm sSFR/yr^{-1}}]$} &
\colhead{$N$\tablenotemark{a}} &
\colhead{$\rho_{\rm Spear}$\tablenotemark{b}} &
\colhead{$\sigma_{\rm Spear}$\tablenotemark{c}} &
\colhead{Slope} &
\colhead{Intercept} &
\colhead{$\sigma_{\rm RMS}$\tablenotemark{d}} &
\colhead{$\sigma_{\rm int}$\tablenotemark{e}}}
\startdata
$\beta_{\rm phot}$ & -9.60...-8.84 & 87 & 0.33 & 3.12 & $0.57\pm0.15$ & $-1.42\pm0.02$ & 0.52 & 0.33 \\
\nodata & -8.84...-8.00 & 87 & 0.56 & 5.24 & $0.73\pm0.11$ & $-1.67\pm0.03$ & 0.38 & 0.21 \\
\\
$\beta_{\rm SED}$ & -9.60...-8.84 & 87 & 0.55 & 4.42 & $0.95\pm0.14$ & $-1.48\pm0.02$ & 0.47 & 0.31 \\
\nodata & -8.84...-8.00 & 87 & 0.65 & 6.07 & $0.87\pm0.09$ & $-1.78\pm0.03$ & 0.33 & 0.20 \\
\enddata
\tablenotetext{a}{Number of galaxies.}
\tablenotetext{b}{Spearman rank correlation coefficient.}
\tablenotetext{c}{Standard deviations from null hypothesis.}
\tablenotetext{d}{Total RMS.}
\tablenotetext{e}{Intrinsic dispersion (see text).}
\label{tab:linfit}
\end{deluxetable*}

\subsubsection{Contribution of A Stars to the Near-UV Flux}
\label{sec:astars}

\begin{figure}[t]
\plotone{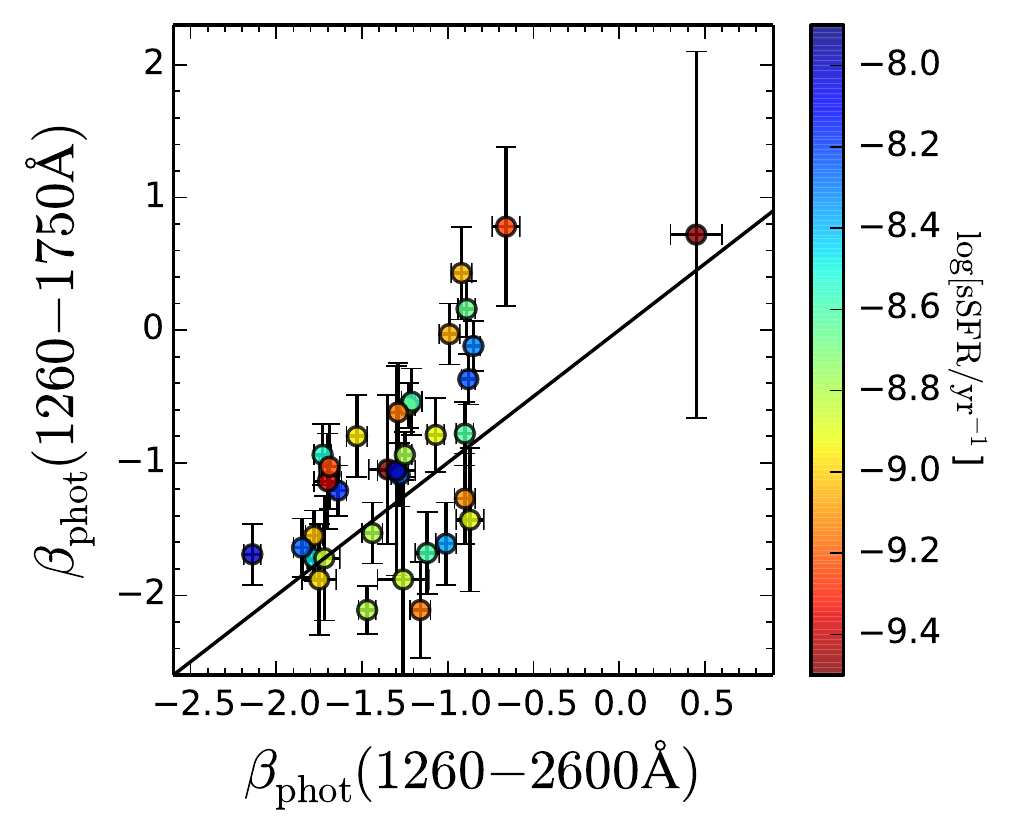}
\caption{Comparison of UV spectral slopes computed over different
  wavelength ranges that include ($\beta_{\rm phot}(1260-2600\,{\rm
    \AA})$) and exclude ($\beta_{\rm phot}(1260-1750\,{\rm \AA})$) the
  near-UV portion of the spectrum that can have a significant
  contribution from stars of A2 and later spectral type.  We find no
  evidence that $\beta_{\rm phot}(1260-1750\,{\rm \AA})$ is
  systematically bluer than $\beta_{\rm phot}(1260-2600\,{\rm \AA})$,
  suggesting that A stars do not contribute significantly to the
  near-UV continuum, even for galaxies with lower sSFRs.}
\label{fig:betabluecompare}
\end{figure}

We have shown that galaxies with lower sSFRs have on average redder UV
slopes at a given Balmer optical depth, but it remains to be seen
whether late-type stars are responsible for this trend.  To assess the
contribution of lower mass A stars to the UV continuum, we
recalculated the UV slopes using only those photometric points that
lie blueward of rest-frame $1750$\,\AA, $\beta(1260-1750\,{\rm \AA})$.
Only O and B stars, and the most massive A stars (spectral type A1,
A0) contribute significantly at these blue wavelengths (e.g.,
\citealt{leitherer95, calzetti94}).  We compared these ``blue'' UV
slopes to those obtained using photometry over the full wavelength
range between $1260$ and $2600$\,\AA, $\beta(1260-2600\,{\rm \AA})$.
The comparison (Figure~\ref{fig:betabluecompare}) includes only those
galaxies that had a minimum of 4 photometric points available for
calculating $\beta$ when excluding all points with $\lambda >
1750$\,\AA.\footnote{The galaxies shown in
  Figure~\ref{fig:betabluecompare} are all in the COSMOS field, where
  $\beta$ is determined using both broad- and intermediate-band
  photometry with sufficient independent wavelength sampling to
  accurately compute the UV slope in the relevant wavelength ranges.}

We find no evidence that galaxies with lower sSFRs have $\beta_{\rm
  phot}(1260-1750\,{\rm \AA})$ bluer than $\beta_{\rm
  phot}(1260-2600\,{\rm \AA})$, as would be expected if the near-UV
continua of such galaxies were weighted more heavily by the flux from
A stars.  The same conclusion is reached when considering the relation
between $\beta_{\rm phot}(1260-1750\,{\rm \AA})$ and the slope
obtained using photometry strictly in the near UV, $\beta_{\rm
  phot}(1750-2600\,{\rm \AA})$.

The redness of the UV slope measured over the shorter baseline in
wavelength can in principle reflect the more severe attenuation at
these short wavelengths, an issue that we discuss briefly in
Section~\ref{sec:betamods}.  Because the difference between $\beta$
measured across the short and long wavelength baselines does not
appear to correlate with sSFR (Figure~\ref{fig:betabluecompare}), the
contribution of older stars to the near-UV continuum is likely minimal
over the range of sSFR considered here, and thus not a dominant factor
in the spread or trend in $\beta$ with $\tau_{\rm b}$.

\subsection{Effect of Metallicity on the UV Slope}

An issue separate from the effect of older stars on the near-UV
continuum is the extent to which differences in metallicities of the
stellar populations lead to trends between $\beta$ and $\tau_{\rm b}$.
We investigated the dependence of $\beta$ on metallicity by using
MOSDEF measurements of the [NII]$\lambda6585$/H$\alpha$ ratio (``N2
index'') with the \citet{pettini04} calibration to infer the gas-phase
oxygen abundance.\footnote{A discussion of the different metallicity
  indicators accessible with MOSDEF are discussed in
  \citet{sanders15}.  Assuming these different indicators does not
  alter our conclusions regarding the change in $\beta$ with inferred
  oxygen abundance.}  We have assumed that the gas-phase and stellar
metallicities track each other closely, as should be the case for the
short-lived O and B stars that dominate the UV continuum.  The range
of metallicities probed by our sample is $8.1\la 12+\log(O/H)\la 8.7$
(i.e., $0.2\la Z\la 1.0$\,Z$_\odot$, assuming the value of the solar
abundance from \citet{allende01} and \citet{asplund04}), with a mean
oxygen abundance that changes by $\simeq 0.3$\,dex between the reddest
and bluest UV slopes.  For a \citet{chabrier03} IMF and a constant
star formation for $100$\,Myr, the \citet{bruzual03} stellar
population synthesis models imply a difference in intrinsic UV slope
of $\Delta\beta_{\rm int}\simeq 0.2$ for the range of metallicites
probed in our sample.  This difference is much smaller than the spread
in $\beta$ at a given $\tau_{\rm b}$, and much smaller than the change
in $\beta$ as $\tau_{\rm b}$ increases
(Figure~\ref{fig:beta_vs_taub}).  We conclude that differences in the
metallicities of the stellar populations are unlikely to be the
dominant factor in driving the change in UV slope with increasing
Balmer line opacity.

\subsection{Scatter in $\beta$ in the limit where $\tau_{\rm b} \rightarrow 0$}
\label{sec:betavstauscatter}

The previous discussion of the effects of late-type stars and
metallicity on the UV slope leads to the conclusion that dust
obscuration is likely the underlying cause of the trend between
$\beta$ and Balmer optical depth.  In this scheme, variations in the
dust opacity between individual galaxies are likely responsible for
the scatter in $\beta$ at a fixed $\tau_{\rm b}$.  Such variations in
the opacity may be due to a dependence on dust composition, resulting
in different absorption and scattering cross-sections of the dust
grains.  Alternatively, the geometrical distribution of the dust with
respect to the ionizing stars in galaxies can lead to variations in
dust opacity.  Galaxies of a given $\tau_{\rm b}$ may exhibit a wide
range of $\beta$ depending on the dust column densities towards the
ionized regions and the stellar continuum (e.g., \citealt{calzetti94,
  calzetti00}).  Though, as emphasized in \citet{calzetti94}, the
spatial distribution of dust and stars cannot explain the large
scatter in $\beta$ in the limit where $\tau_{\rm b}\rightarrow 0$.  In
this limit, we expect $\beta$ to approach its intrinsic value,
typically between $\beta_{\rm int}\simeq -2.6$ and $-2.0$, depending
on the details (e.g., IMF, metallicity) of the stellar population
model.

There are several reasons to suspect that galaxies with $\tau_{\rm
  b}=0$ are not dust-free, even though their Balmer decrements would
indicate formally otherwise.  First, such galaxies have measurement
errors in $\tau_{\rm b}$ that do not exclude the presence of some
dust.  Second, even galaxies with the bluest UV spectral indices in
our sample have metallicities of at least $12+\log(O/H)\ga 8.1$
($\approx 0.26$\,Z$_\odot$) implying that they must contain some dust.
Third, if we de-redden the best-fit SEDs assuming a \citet{calzetti00}
attenuation curve and the best-fit $E(B-V)$, we obtain $\beta_{\rm
  int}\simeq -2.3$.  This is significantly bluer than the $\beta_{\rm
  int} \approx -1.3$ and $-1.6$ implied by the linear regression for
the bin of lowest sSFR (Table~\ref{tab:linfit}).  Moreover, the
detection of H$\alpha$ emission in these galaxies requires the
presence of hydrogen-ionizing stars, and the blue hot stars will cause
an otherwise intrinsically shallow UV spectral index to become steep
(i.e., more negative).  Finally, from an observational standpoint, if
the rest-UV emission in these galaxies is not spatially coincident
with the rest-optical emission, then the observed Balmer decrement may
not be indicative of the global reddening of the galaxy as deduced
from the UV slope, a case that may apply for all galaxies in our
sample.  For these reasons, the linear regressions between $\beta$ and
$\tau_{\rm b}$ should be interpreted with caution in the limit
$\tau_{\rm b} \rightarrow 0$.

To summarize, the results of Section~\ref{sec:variations_ssfr} (e.g.,
Figures~\ref{fig:beta_vs_taub} and \ref{fig:linfit}) imply that the
physical origin of the scatter in $\beta$ vs. $\tau_{\rm b}$ must vary
systematically with sSFR, a point that we discuss further in the next
section.  We have investigated the potential effect of A stars on our
measurements of $\beta$ and find that it cannot explain the scatter in
$\beta$ at a given $\tau_{\rm b}$.  Further, variations of the
high-mass end of the IMF (e.g., as a function of metallicity) lead to
corrections of the UV slope of $\approx 10\%$ \citep{calzetti94},
which again cannot account for the large scatter in $\beta$ at a given
$\tau_{\rm b}$.  In Section~\ref{sec:discussion}, we put forth a
physical interpretation that is consistent with the trends and scatter
between $\beta$ and $\tau_{\rm b}$.

\subsection{The Attenuation Curve of High-Redshift Galaxies}
\label{sec:attcurve}

Given evidence that neither an older stellar population nor a change
in metallicity is responsible for the reddening of $\beta$ with
increasing $\tau_{\rm b}$, we are left with the possibility that dust
attenuation is the dominant factor in the observed trend between
$\beta$ and $\tau_{\rm b}$.  A similar line of reasoning is presented
in \citet{calzetti94} to account for the correlation between UV slope
and Balmer optical depth for a sample of 39 local starburst and blue
compact galaxies from \citet{kinney93}, where the local trend is shown
in Figure~\ref{fig:linfit}.  Interestingly, we find that the
correlation between UV slope and Balmer optical depth is steeper for
the local sample, relative to that observed for the $1.4\la z\la 2.6$
MOSDEF sample.  The median sSFR of the local galaxies is $\log({\rm
  sSFR}/{\rm yr}^{-1}) \approx -9.00$, and these galaxies generally
have both lower SFRs ($\simeq 4$\,$M_\odot$\,yr$^{-1}$) and lower
stellar masses ($\simeq 3.9\times 10^{9}$\,$M_\odot$) compared to
those in our sample.  A more detailed comparison of our results with
local samples (including the SDSS) will be presented elsewhere.  For
the moment, we note that the segregation of the $\beta$ vs. $\tau_{\rm
  b}$ relations (Figure~\ref{fig:linfit}) indicates an sSFR-dependent
scaling between nebular and stellar attenuation in high-redshift
galaxies that differs from local ones (Section~\ref{sec:nebvsstel}).

\subsubsection{Derivation of the Selective Attenuation Curve}
\label{sec:attder}

\begin{figure}
\plotone{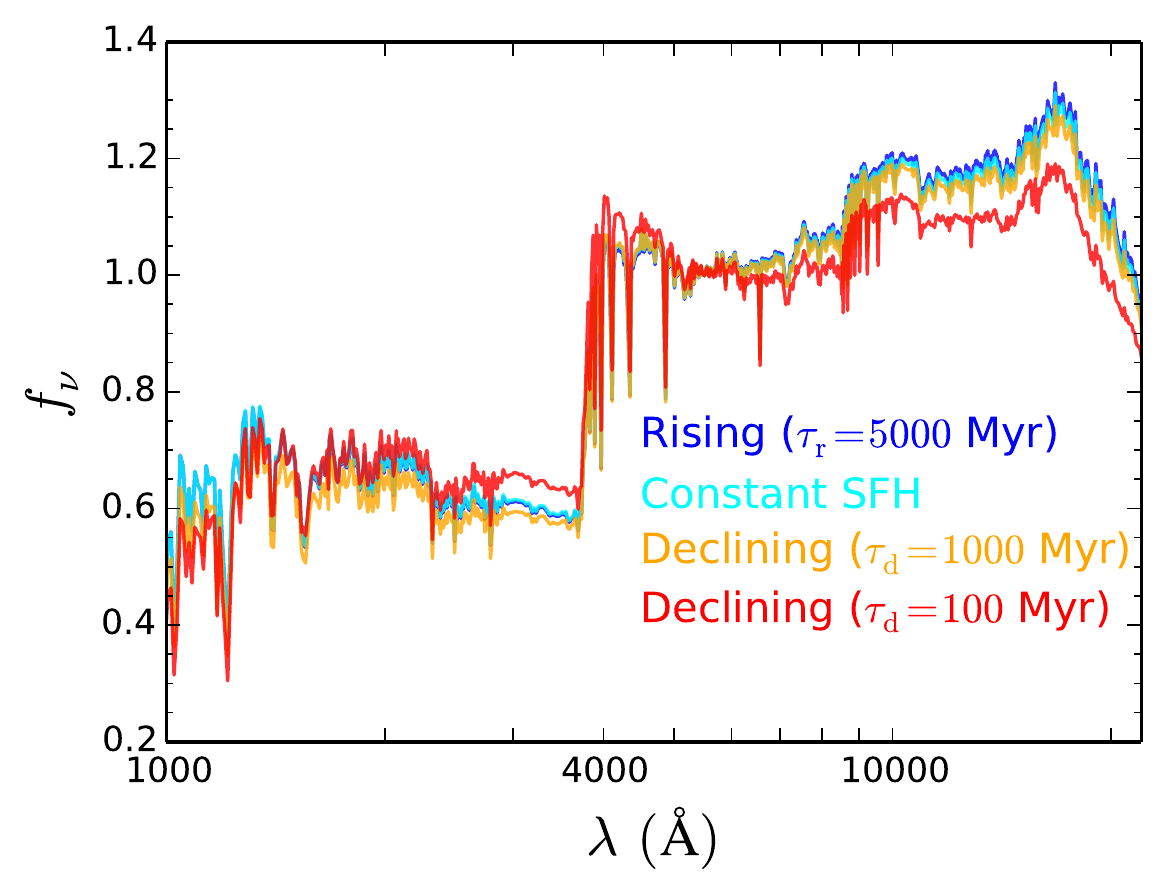}
\caption{Stellar SEDs for a galaxy with a stellar mass of $M^{\ast}
  \simeq 1.2\times 10^{10}$\,$M_\odot$ and SFR$\simeq
  16$\,$M_\odot$\,yr$^{-1}$, assuming different simple star-formation
  histories: an exponentially rising model with an e-folding time of
  5000\,Myr, constant star formation, and exponentially declining
  models with e-folding times of 1000 and 100\,Myr.  The corresponding
  ages of the stellar populations are $\approx$ 800, 3000, 720, 570,
  and 200\,Myr, respectively.  The SEDs are cast in terms of $f_\nu$
  and are normalized so that $f_\nu(\lambda=5500\,{\rm \AA})=1$.
  Fixing the sSFR of the galaxy results in similar SED shapes for most
  of the assumed star-formation histories.}
\label{fig:conformity}
\end{figure}

In what follows, we adopt a procedure similar to that presented in
\citet{calzetti94}, generalized from that of \citet{kinney94}, to
derive the selective attenuation curve.  The premise behind this
procedure is that because the UV spectral index is sensitive primarily
to dust attenuation, the ratio of the average spectrum of galaxies
with high Balmer optical depths to that of galaxies with low optical
depths should reflect the average effect of attenuation on the stellar
SEDs.  Of course, especially in the optical and near-IR, there can be
a wide range of intrinsic SED shapes for star-forming galaxies
depending on their star-formation histories and ages.  However, to
first order, galaxies of a given sSFR will have comparably-shaped
intrinsic stellar SEDs, as the flux at any given wavelength is
contributed to in a similar proportion by the current star formation
and the light from previous generations of star formation.  The exact
form of how the previous stellar mass was built up (i.e., the
star-formation history) is unimportant, as the rest-frame optical to
near-IR light that is used to infer stellar masses (and hence sSFRs)
arises from some combination of the {\em currently} forming stellar
population, and from lower mass (spectral type F5 and later) stars
that have not yet evolved off the main sequence for the age of the
Universe at $z\sim 2$.\footnote{This assumes an IMF that does not
  evolve with redshift above $z\sim 2$.  Some circumstantial evidence
  for an un-evolving IMF at $z\ga 2$ comes from the general agreement
  between the integral of the cosmic SFR density and the evolution of
  the stellar mass density \citep{reddy09, madau14}.}  For instance,
Figure~\ref{fig:conformity} shows the conformity of the stellar SED
for a galaxy with an SFR and $M^{\ast}$ (and hence, sSFR) typical of
the average galaxy in our sample, irrespective of the star-formation
history.  SEDs that assume a relatively short decay timescale for star
formation (e.g., $\tau_{\rm d}\la 100$\,Myr) deviate by $\approx 10\%$
at optical and near-IR wavelengths, but such histories are ruled out
for the vast majority of objects in our sample given their typical
stellar population ages of $\ga 300$\,Myr, combined with the evidence
that, at least on average, $>L^{\ast}$ galaxies at $z\sim 2$ show
evidence for rising star formation at previous epochs (e.g.,
\citealt{reddy12b}).

\begin{figure}
\plotone{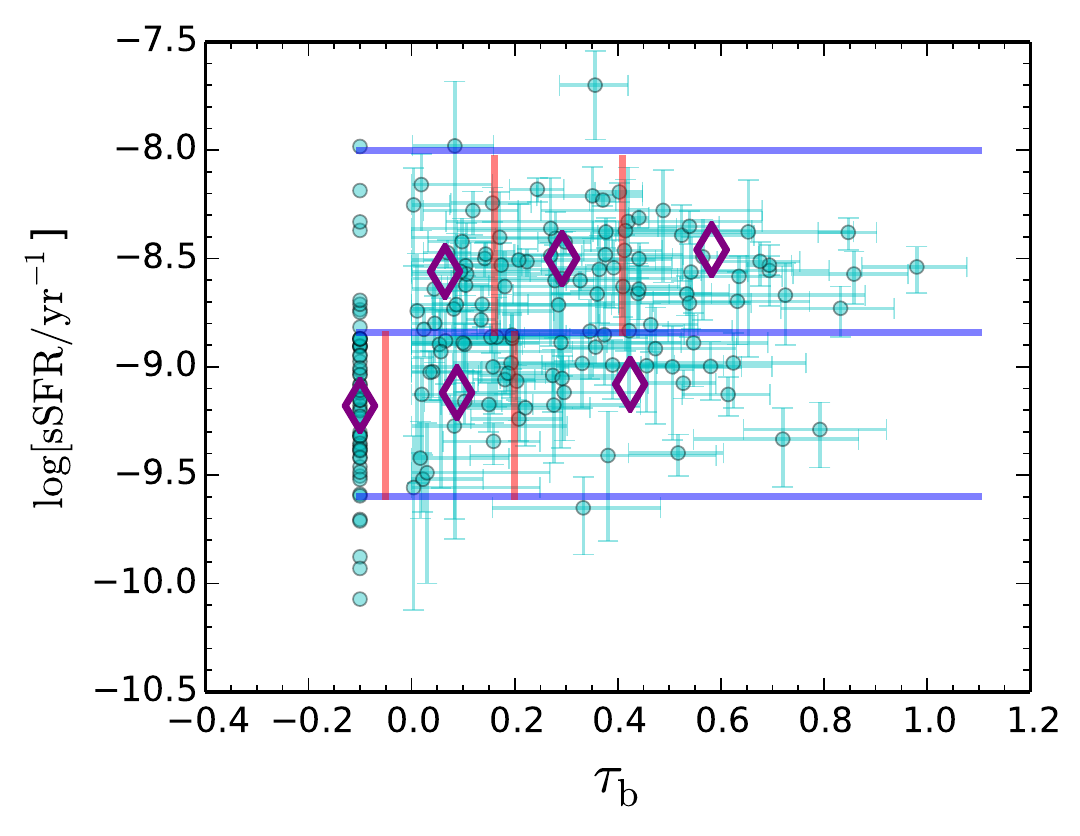}
\caption{Distribution of sSFR with Balmer optical depth.  Objects with
  $\tau_{\rm b}=0$ are offset to the left for clarity.  The horizontal
  lines denote the boundaries of the two sSFR bins and the vertical
  lines denote the boundaries of the three $\tau_{\rm b}$ bins per
  sSFR bin used to calculate the attenuation curve.  The open diamonds
  indicate the mean values of $\log[{\rm sSFR/yr^{-1}}]$ and
  $\tau_{\rm b}$ for each bin.}
\label{fig:ssfrvstau}
\end{figure}

Motivated by the apparent sSFR-dependence of the $\beta$
vs. $\tau_{\rm b}$ relation, we derived the attenuation curve for
galaxies in two bins of sSFR.  We first examined the distribution of
sSFR with $\tau_{\rm b}$ (Figure~\ref{fig:ssfrvstau}).  Approximately
96$\%$ of the galaxies in our sample (174 of 182) lie in the range
$-9.6\le\log[{\rm sSFR/yr^{-1}}]<-8.0$.  Dividing the number of
objects in this range into two bins, with equal numbers of objects
(87) per bin, results in the following sSFR bins: $-9.60\le\log[{\rm
    sSFR/yr^{-1}}]<-8.84$ (sSFR bin 1) and $-8.84\le\log[{\rm
    sSFR/yr^{-1}}]<-8.00$ (sSFR bin 2).  The objects within each sSFR
bin were further divided into three bins of $\tau_{\rm b}$.  For sSFR
bin 1, the first $\tau_{\rm b}$ bin includes all objects with
$\tau_{\rm b}=0$, and the remaining objects are split roughly equally
into two additional bins.  Given the low number of objects (8) that
have $\tau_{\rm b}=0$ in sSFR bin 2, for this sSFR bin we divided the
objects equally into three bins of $\tau_{\rm b}$.

The derivation of the attenuation curve relies on the assumption that
the {\em average} intrinsic SED (and hence the sSFR) is unchanging as
a function of Balmer optical depth, irrespective of the exact mix of
different star-formation histories that may be contributing to that
average SED.  Figure~\ref{fig:ssfrvstau} shows that the average sSFRs
of galaxies in each bin of $\tau_{\rm b}$ is roughly constant,
implying that a simple ratio of the average SEDs obtained in each bin
will directly probe the wavelength dependence of dust obscuration.  In
Section~\ref{sec:attunc}, we address the slight ($<0.1$\,dex)
differences in the $\langle \log[{\rm sSFR/yr^{-1}}]\rangle$ obtained
in each bin, and how these differences translate into a systematic
uncertainty in our derived attenuation curve.

\begin{figure*}
\plotone{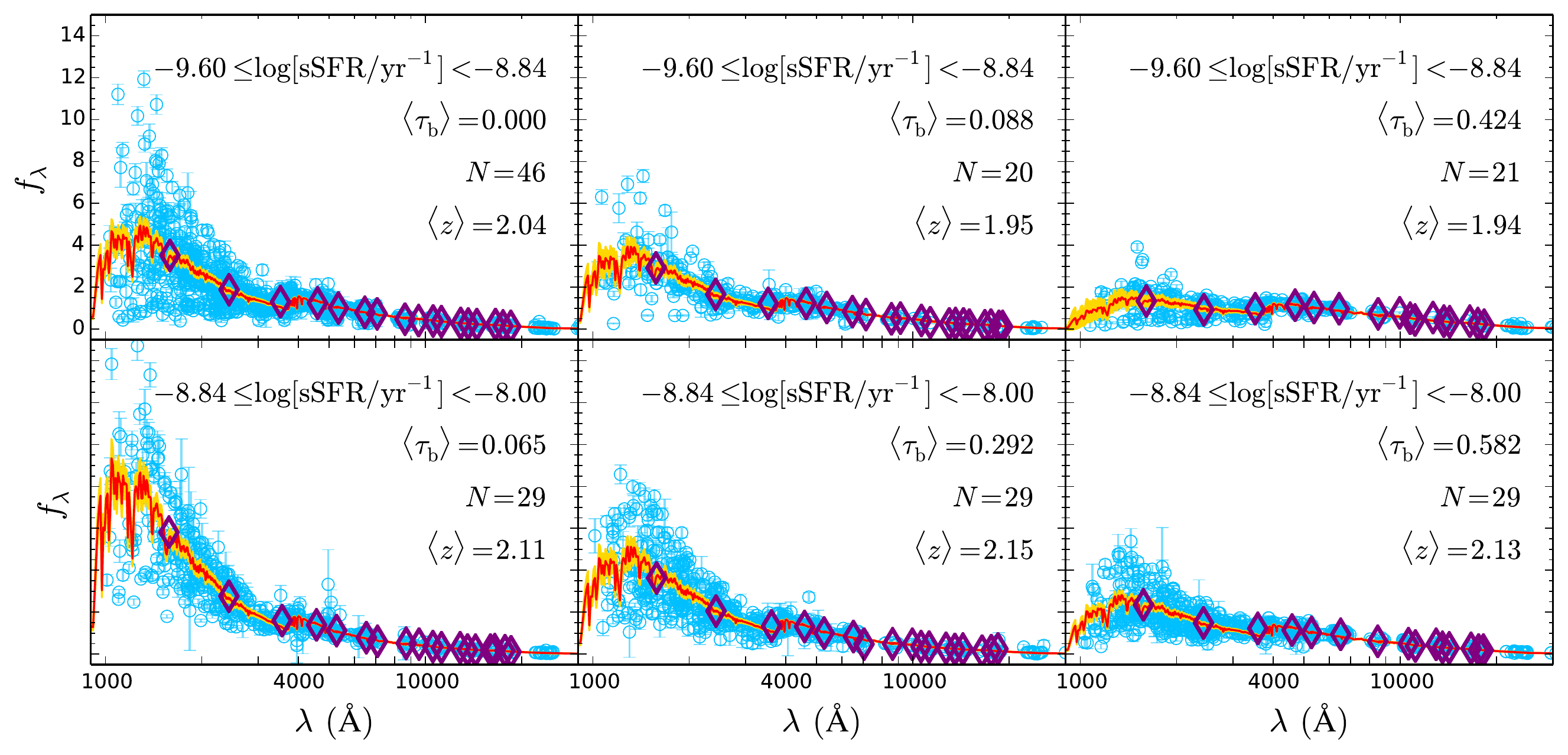}
\caption{Photometry ({\em circles}) and model fits ({\em red}) of
  objects in bins of sSFR and Balmer optical depth, with number of
  contributing objects and mean redshifts indicated.  The photometry
  has been normalized such that $f_{\lambda}(5500\,{\rm \AA}) = 1$.
  The fitting takes into account the photometric (flux) errors and
  allows for a range of possible star-formation histories, ages, color
  excesses, and attenuation curves (see text).  The orange shaded
  regions indicate the $1$\,$\sigma$ dispersion about the mean SED
  fits, based on Monte Carlo simulations where the sSFRs, $\tau_{\rm
    b}$, and photometric errors for each object are perturbed and the
  mean SED is re-calculated.  For comparison, the diamonds show the
  un-weighted average of the fluxes in 1000\,\AA-width bins in
  wavelength.}
\label{fig:aveseds}
\end{figure*}

Having established the aforementioned bin definitions, we then
estimated the average SED of objects in each bin as follows.  The
broad- and intermediate-band fluxes measured for each object were
normalized by the flux of the best-fit SED for that object at
rest-frame $\lambda=5500$\,\AA\, and then shifted to the rest frame
based on its spectroscopic redshift.  The normalized flux density
measurements for all objects were then combined and fit with a
\citet{bruzual03} stellar population model.  In the fitting, we
assumed an equal weighting for all objects irrespective of the flux
measurement errors so that no single object dominated the fit, though
we note that taking into account photometric errors results in fits
that are similar to those obtained here.  To ensure the best possible
fit to the data, we allowed for a range of possible star-formation
histories (declining, constant, and rising), no restriction on the age
of the model apart from it being less than the age of the Universe at
the mean redshift of the sample $\langle z\rangle \sim 2.0$, color
excesses from $E(B-V)=0.00$ to $0.60$, and four attenuation curves, a
Calzetti starburst reddening curve \citep{calzetti00}, and the
line-of-sight SMC, LMC, and Galactic extinction curves.  The exact
details of the best-fit model are not essential, so long as the model
provides a good fit to the fluxes of the contributing objects.
Figure~\ref{fig:aveseds} shows the best-fit functional form to the
normalized photometry of all objects in each bin of sSFR and
$\tau_{\rm b}$.  In this regard, also shown are the weighted mean
fluxes of objects in $1000$\,\AA-width bins in wavelength (i.e., the
empirical SED), demonstrating that the model fits are not in any way
skewed relative to these mean empirical values.  Below, we also
demonstrate that our model fits result in a shape of the attenuation
curve that follows the one calculated from the empirical SEDs.

The random uncertainty in the average SEDs was estimated as follows.
For each object, we perturbed its sSFR and $\tau_{\rm b}$ based on the
errors in these quantities, and then rebinned all the objects
according to these perturbed values.  The photometry of each object
was also perturbed according to the photometric errors, and the
average SEDs were recalculated.  This procedure was repeated 100 times
to obtain statistically robust determinations of the standard
deviations in the average SEDs, which are shown by the shaded regions
in Figure~\ref{fig:aveseds}.

\begin{figure*}
\plotone{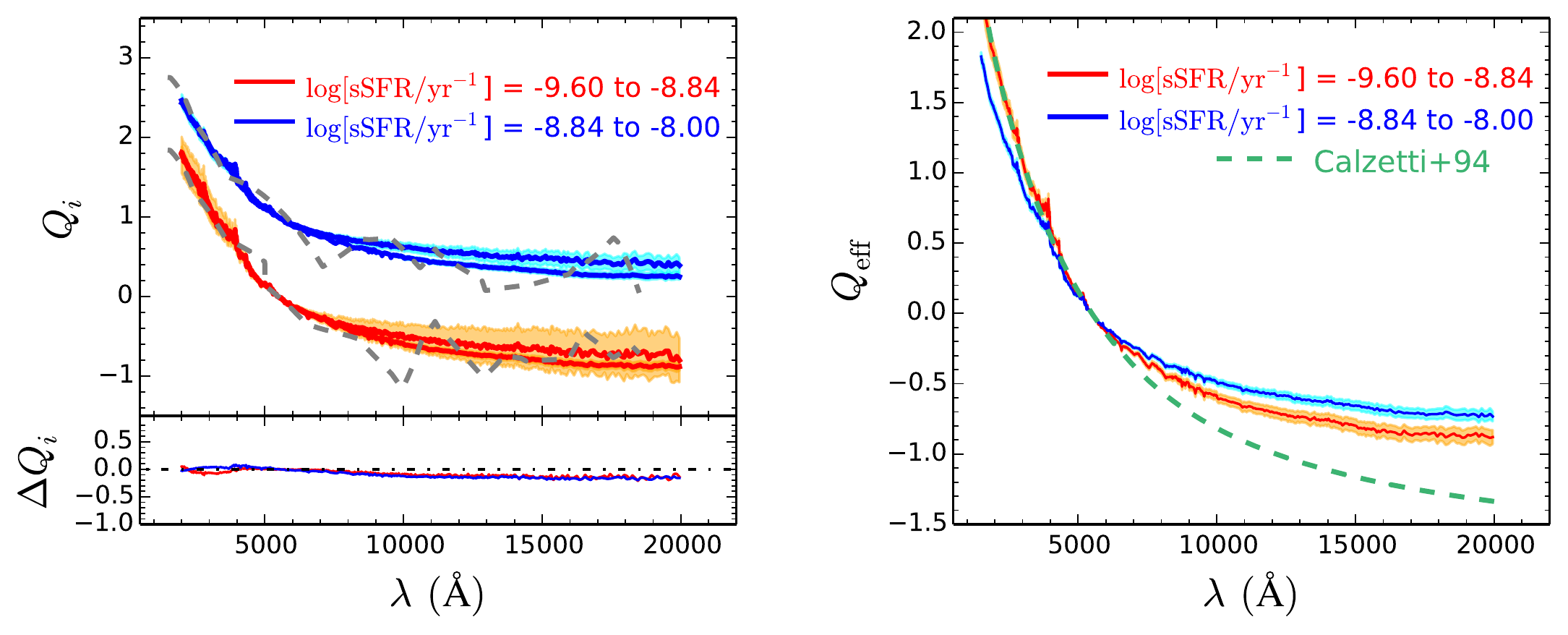}
\caption{ {\em Left:} Selective attenuation curves ($Q_2$ and $Q_3$)
  calculated using objects in two bins of sSFR.  The curves for sSFR
  bin 2 have been shifted upward by $\Delta Q_i=1$ for clarity.  The
  shaded regions indicate the uncertainties in $Q_i$.  The $Q_i$
  inferred from the empirical SEDs are denoted by the dashed lines.
  Also shown at the bottom is the difference between $Q_3$ and $Q_2$
  as a function of wavelength for each bin of sSFR.  {\em Right:}
  Effective attenuation curves ($Q_{\rm eff}$) calculated as the
  weighted average of the $Q_2$ and $Q_3$ shown in the {\em left}
  panel for each sSFR bin.  The weighted uncertainties in $Q_{\rm
    eff}$ are indicated by the shaded regions.  The dashed line shows
  the \citet{calzetti94} selective attenuation curve.  }
\label{fig:selq}
\end{figure*}

The {\em selective attenuation} between each of these average
(composite) SEDs is then defined as
\begin{equation}
Q_i (\lambda) \equiv -\frac{1}{\tau_i -
  \tau_1}\ln\left[\frac{F_i(\lambda)}{F_1(\lambda)}\right],
\label{eq:qeff}
\end{equation}
where $\tau_i$ and $F_i$ refer to the average Balmer optical depth and
the composite spectrum, respectively, of the objects in the $i$th bin
of $\tau_{\rm b}$ \citep{calzetti94}.  $Q_i$ is computed relative to
the first bin where by construction $\tau_1=0$ for the first sSFR bin,
and $\tau_1 = 0.065$ for the second sSFR bin.  Thus, for each bin of
sSFR, we obtain two estimates of the selective attenuation ($Q_2$ and
$Q_3$).  Figure~\ref{fig:selq} shows our calculation of $Q_2$ and
$Q_3$ for the two bins of sSFR specified above.  The uncertainties in
$Q_i$ were determined by randomly perturbing the composite SEDs
according to their respective uncertainties and recomputing $Q_i$ for
these realizations.  The errors in the mean $Q_i$ are indicated by the
shaded regions in Figure~\ref{fig:selq}, and are $\sigma(Q_2) \approx
0.20$ and $\sigma(Q_3) \approx 0.05$ for sSFR bin 1, and $\sigma(Q_2)
\approx 0.10$ and $\sigma(Q_3) \approx 0.04$ for sSFR bin 2.  The
differences between $Q_2$ and $Q_3$ for a given sSFR bin are $\Delta
Q_i \la 0.15$, or $\la 0.07$\,dex, and are of the same order as the
random uncertainties in the mean $Q_i$.

Additionally, we calculated the $Q_i$ based on the empirically
determined SEDs.  This was accomplished by linearly interpolating
between the average flux points of each composite SED (i.e., as shown
by the diamonds in Figure~\ref{fig:aveseds}) to produce an empirical
SED.  These empirical SEDs were then combined in the same way as the
model fits in order to calculate $Q_i$ (i.e., using
Equation~\ref{eq:qeff}), and the results are denoted by the dashed
lines in the left panel of Figure~\ref{fig:selq} (for clarity, we show
the results for $Q_3$).  The empirically determined $Q_i$ track
closely those inferred from our model fits, and for our subsequent
analysis, we adopt the latter.

Our analysis indicates that the $Q_i$ within a given sSFR bin are very
similar in shape, implying that the average SEDs used to compute $Q_i$
have roughly similar stellar populations (as would be expected given
the discussion above regarding the similarity in SED shapes for a
given sSFR).  In other words, the shape of $Q_i$ is dictated primarily
by differences in the dust obscuration, as parameterized by $\tau_{\rm
  b}$, between the average SEDs show in Figure~\ref{fig:aveseds}, and
not by significant differences in the average intrinsic stellar
populations from one $\tau_{\rm b}$ bin to another (see
Section~\ref{sec:attunc} for a full discussion of the systematic
uncertainties in the attenuation curve).  Given the similarity in the
shapes of $Q_2$ and $Q_3$ for each sSFR bin, we simply took a weighted
average of the two selective attenuation curves to produce what we
refer to as the ``effective'' attenuation curve, $Q_{\rm eff}$.  The
weights were calculated as $1/\sigma_\lambda^2$, where
$\sigma_\lambda$ is the uncertainty in $Q_i$ as a function of
wavelength.  The uncertainty in the weighted average is $\sigma_{\rm
  weighted, \lambda} = \sqrt{1/\sum \sigma_\lambda^{-2}}$.

We fit the effective attenuation curve in each sSFR bin using
third-order polynomial functions of $1/\lambda$ over the wavelength
ranges $\lambda=0.15-0.60$\,$\mu$m and $\lambda=0.60-2.20$\,$\mu$m.
These specific ranges were chosen to ensure the best possible
representation of the data using polynomials of the smallest order.
The polynomial fit for sSFR bin 1 ($-9.60\le\log[{\rm
    sSFR/yr^{-1}}]<-8.84$) is:
\begin{eqnarray}
& & Q_{\rm eff}(\lambda,-9.60\le\log[{\rm sSFR/yr^{-1}}]<-8.84 ) \nonumber \\
& = & -2.101 + 1.450/\lambda - 0.181/\lambda^2 + 0.010/\lambda^3, \nonumber \\
                    &   & \,\,\,\,\,0.15\le \lambda < 0.60\,\mu{\rm m}; \nonumber \\
                    & = & -0.955 - 0.169/\lambda + 0.732/\lambda^2 - 0.200/\lambda^3, \nonumber \\
                    &   & \,\,\,\,\,0.60\le \lambda < 2.20\,\mu{\rm m}.
\label{eq:qeff1_fit}
\end{eqnarray}
The polynomial fit for sSFR bin 2 ($-8.84\le\log[{\rm sSFR/yr^{-1}}]<-8.00$)
is:
\begin{eqnarray}
& & Q_{\rm eff}(\lambda,-8.84\le\log[{\rm sSFR/yr^{-1}}]<-8.00) \nonumber \\
& = & -1.834 + 1.299/\lambda - 0.179/\lambda^2 + 0.010/\lambda^3, \nonumber \\
                    &   & \,\,\,\,\,0.15\le \lambda < 0.60\,\mu{\rm m}; \nonumber \\
                    & = & -0.878 + 0.136/\lambda + 0.348/\lambda^2 - 0.091/\lambda^3, \nonumber \\
                    &   & \,\,\,\,\,0.60\le \lambda < 2.20\,\mu{\rm m}.
\label{eq:qeff2_fit}
\end{eqnarray}
The error in $Q_{\rm eff}$ at each wavelength point was used to weight
the data appropriately in the fitting.  The formal uncertainty in the
model fit at each wavelength varies in the range $\sigma_{\rm
  fit}\approx 0.001$ to $0.004$.  As discussed below, several
additional steps are required to convert $Q_{\rm eff}$ to a {\em
  total} attenuation curve, the reason being that the former is
computed using galaxies of different luminosities at different
redshifts, and therefore the normalization of the attenuation curve
must be determined separately.  Nonetheless, $Q_{\rm eff}$ is useful
in two respects: it indicates the wavelength dependence of the
obscuration of the stellar continuum and the optical depth to dust of
the continuum relative to the Balmer lines.  We discuss these points
below.

\subsubsection{Comparison of the Average Optical Depths of the Continuum
and the Balmer Lines}
\label{sec:nebvsstel}

The definition of $Q_{\rm eff}$ (Equation~\ref{eq:qeff}) implies that
the Balmer optical depth, $\tau_{\rm b}$, is related to the difference
in optical depths of the continuum at the wavelengths of H$\beta$ and
H$\alpha$: $\tau_{\rm c}({\rm H\beta})$-$\tau_{\rm c}({\rm H\alpha}) =
[Q_{\rm eff}({\rm H\beta})- Q_{\rm eff}({\rm H\alpha})] \tau_{\rm b}$
(e.g., \citealt{calzetti94}), assuming that $\langle \tau_{\rm
  b}\rangle =0$ for the first bin of $\tau_{\rm b}$.  According to
Equations~\ref{eq:qeff1_fit} and \ref{eq:qeff2_fit}, we can then write
$\tau_{\rm c}({\rm H\beta})-\tau_{\rm c}({\rm H\alpha}) = (0.42\pm
0.01)\tau_{\rm b}$ and $\approx (0.36\pm 0.01)\tau_{\rm b}$ for sSFR
bins $-9.60\le\log[{\rm sSFR/yr^{-1}}]<-8.84$ and $-8.84\le\log[{\rm
    sSFR/yr^{-1}}]<-8.00$, respectively.\footnote{The relation is only
  approximate for bin where $-8.84\le\log[{\rm sSFR/yr^{-1}}]<-8.00$
  as the average Balmer optical depth for the bin of lowest $\tau_{\rm
    b}$ is not exactly zero.}  Thus, we conclude that analogous to
local galaxies (e.g., \citealt{fanelli88, calzetti94, mashesse99,
  kreckel13}), the ionized gas in high-redshift galaxies with stellar
masses $\ga 10^{9}$\,$M_\odot$ is subject to a higher dust optical
depth {\em on average} than the stellar continuum.  Qualitatively,
this result is consistent with the average difference in color excess
between the gas and stars observed in local star-forming galaxies
\citep{calzetti00, kreckel13}.  We return to this point in
Section~\ref{sec:gasvsstars}, where we investigate in detail the
relationship between the total attenuation of the ionized gas and
stellar continuum for individual galaxies.

\subsubsection{Shapes of the Attenuation Curves}

\begin{figure}
\plotone{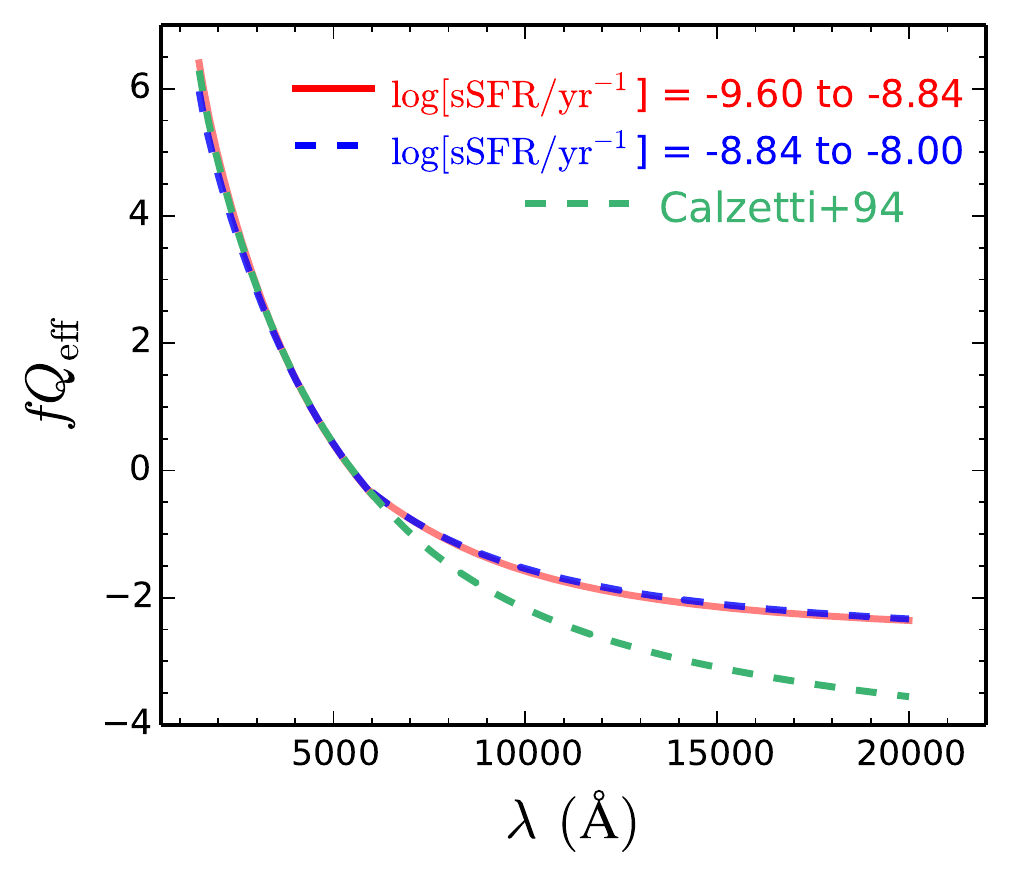}
\caption{$fQ_{\rm eff}$ for the two bins of sSFR as indicated in the
  figure, where the multiplicative factors $f=2.676$ and $3.178$ are
  required to ensure that $f[Q_{\rm eff}(B)-Q_{\rm eff}(V)] =
  k(B)-k(V)=1$ (see text).  The \citet{calzetti94} attenuation curve
  is shown for comparison, where we have applied a similar factor
  $f\approx 2.659$ \citep{calzetti00}.}
\label{fig:fqeff}
\end{figure}

The total attenuation curve is defined as
\begin{eqnarray}
k(\lambda) \equiv \frac{A_\lambda}{E(B-V)},
\label{eq:kdef}
\end{eqnarray}
where $A_\lambda$ is the total magnitude of extinction at wavelength
$\lambda$ and $E(B-V)$ is the color excess of the stellar continuum.
Because $E(B-V) = A(B)-A(V)$, we can write $k(B)-k(V)=1$.  Thus, we
multiplied the effective attenuation curves $Q_{\rm eff}$ by the
factors $f=2.676$ and $3.178$ for sSFR bins 1 and 2, respectively, so
that
\begin{eqnarray}
f[Q_{\rm eff}(B)-Q_{\rm eff}(V)] = k(B)-k(V) = 1.
\label{eq:multeq}
\end{eqnarray}
The factors were computed by evaluating the effective attenuation curves
in the $B$ ($\lambda = 4400$\,\AA) and $V$
($\lambda = 5500$\,\AA) bands.  The resulting $fQ_{\rm eff}$ are shown
for the two bins of sSFR in Figure~\ref{fig:fqeff}.

\begin{figure*}
\epsscale{1.}
\plotone{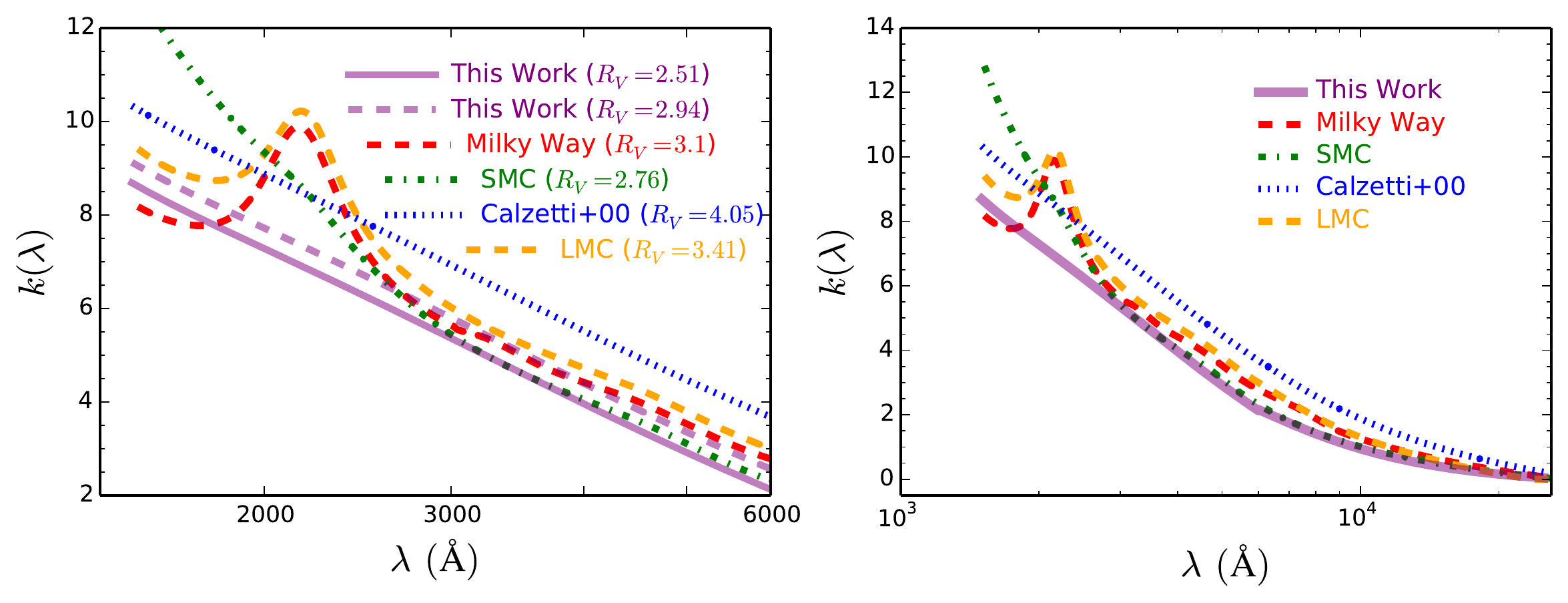}
\caption{{\em Left:} Comparison of the MOSDEF attenuation curve
  assuming two different values of $R_V=2.505$ and $R_V=2.942$ (see
  text), and that of \citet{calzetti00} ({\em dotted} line).  Also
  included are the local extinction curves of the Milky Way, SMC, and
  LMC, where the shapes and normalizations ($R_V$) have been taken
  from \citet{gordon03}. {\em Right:} Same as {\em left} panel, where
  we show the MOSDEF attenuation curve assuming $R_V=2.505$ over the
  full wavelength range $\lambda = 0.15-2.50$\,$\mu$m. }
\label{fig:comparek}
\end{figure*}

The two notable aspects of the effective attenuation curves derived
here are that (a) they appear to deviate at long wavelengths relative
to the \citet{calzetti94} selective attenuation curve, and (b) $Q_{\rm
  eff}$ for the two sSFR bins are strikingly similar.  On the latter
point, {\em a priori} we would not have expected such a good agreement
given the disparity in the trend of UV slope with Balmer optical depth
for galaxies with different sSFRs (Section~\ref{sec:variations_ssfr}).
We return to this issue in Section~\ref{sec:gasvsstars}.  Given the
similarity in $Q_{\rm eff}$ derived for the two sSFR bins, we simply
averaged them, resulting in the following functional form for the {\em total}
attenuation curve:
\begin{eqnarray}
k(\lambda) & = & -5.726 + 4.004/\lambda - 0.525/\lambda^2 + 0.029/\lambda^3 + R_V, \nonumber \\
           &   & \,\,\,\,\,0.15\le \lambda < 0.60\,\mu{\rm m}; \nonumber \\
           & = & -2.672 - 0.010/\lambda + 1.532/\lambda^2 - 0.412/\lambda^3 + R_V, \nonumber \\
           &   & \,\,\,\,\,0.60\le \lambda < 2.20\,\mu{\rm m};
\label{eq:krv}
\end{eqnarray}
where $R_V$ has the usual definition as the ratio of
total-to-selective absorption at $V$-band: $R_V = A_V/E(B-V)$.  The
last step in the derivation of the total attenuation curve is to
calculate $R_V$.  This can be accomplished by extrapolating
$k(\lambda)$ to some wavelength (e.g. $\lambda_n = 2.85$\,$\mu$m) and
assuming that $k(\lambda > \lambda_n)\approx 0$.  This particular
value of $\lambda_n=2.85$\,$\mu$m corresponds to where $k(\lambda)$
for the SMC, LMC, and Milky Way are very close to zero
\citep{gordon03}.  In our case, if we extrapolate
Equation~\ref{eq:krv} to $\lambda_n = 2.85$\,$\mu$m and force
$k(\lambda_n)=0$, then $R_V = 2.505$.

Alternatively, a common second approach is to extrapolate $k(\lambda)$
to infinite wavelength, and set $k(\lambda\rightarrow \infty) = 0$.
Doing so for Equation~\ref{eq:krv} results in $R_V = 2.672$.  More
typically, a $k(\lambda >> 1\,\mu{\rm m}) \propto \lambda^{-1}$
dependence is generally assumed.  In this case, we determined the
functional form of $k(\lambda>2.20\,\mu{\rm m})$ by fitting
Equation~\ref{eq:krv} with a power-law in $1/\lambda$ in the range
$\lambda = 1.9-2.2$\,$\mu$m (this is to ensure that $k(\lambda)$ is a
continuous function across $\lambda=2.2$\,$\mu$m).  With this
long-wavelength extrapolation of $k(\lambda)$, and setting
$k(\lambda\rightarrow \infty) = 0$, we find $R_V = 2.942$.

The different extrapolations discussed above result in a systematic
error in $R_V$ of $\approx 0.4$.  Figure~\ref{fig:comparek} shows the
comparison of the attenuation curves assuming the smallest and largest
values of $R_V$ derived above, along with several other common
extinction and attenuation curves.  Independent estimates of $R_V$
based on mid- and far-infrared data will be presented elsewhere.  For
the remaining analysis, we have assumed the first value of $R_V=2.505$
as it does not require an extrapolation too far beyond where we have
constraints on the shape of the attenuation curve from the photometry
($\lambda \la 2.2$\,$\mu$m).  With this assumption, we arrive at the
final expression for the total attenuation curve relevant for the
stellar continuum:
\begin{eqnarray}
k_{\rm stars}(\lambda) & = & -5.726 + 4.004/\lambda - 0.525/\lambda^2 \nonumber \\
& & + 0.029/\lambda^3 + 2.505, \nonumber \\
& & \,\,\,\,\,0.15\le \lambda< 0.60\,\mu{\rm m}; \nonumber \\
& = &  -2.672 - 0.010/\lambda + 1.532/\lambda^2 \nonumber\\
& & -0.412/\lambda^3 + 2.505, \nonumber \\
& & \,\,\,\,\,0.60\le \lambda<2.85\,\mu{\rm m}. 
\label{eq:kstars}
\end{eqnarray}

\subsubsection{Systematic Uncertainty in $k(\lambda)$ Arising from
Differences in Average Stellar Populations}
\label{sec:attunc}

\begin{figure*}
\epsscale{1.}
\plotone{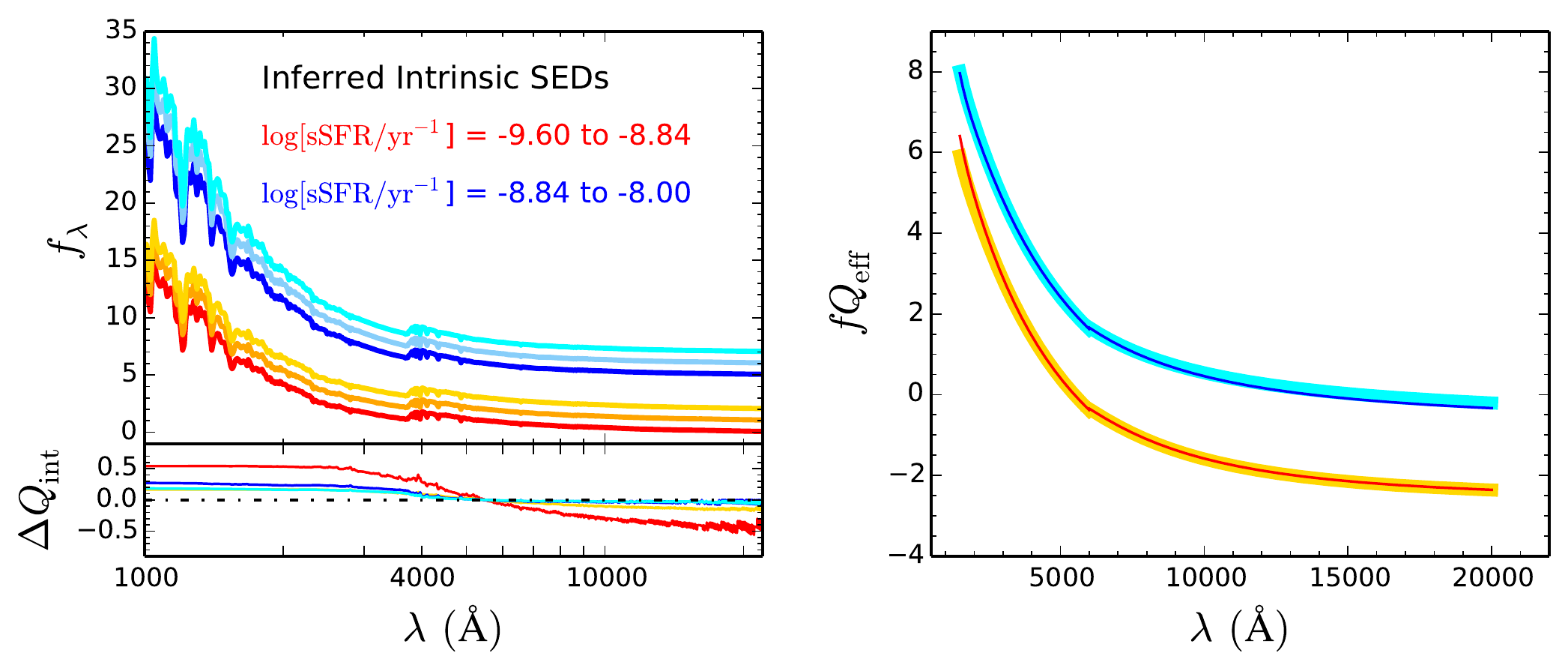}
\caption{{\em Left:} Inferred average intrinsic SEDs for the same bins
  of sSFR and $\tau_{\rm b}$ used to calculate the attenuation curve.
  The intrinsic SED for each galaxy is constructed by assuming a
  simple stellar population with the same SFR(H$\alpha$) and
  $M^{\ast}$ as measured for that galaxy.  These individual SEDs of
  objects in each bin of sSFR and $\tau_{\rm b}$ are then averaged
  together to produce the curves shown in the Figure.  The bottom
  panel shows the implied correction factors to the selective
  attenuation curves due to differences in the intrinsic SEDs of
  galaxies from bin-to-bin.  {\em Right:} Comparison of the $fQ_{\rm
    eff}$ ({\em thin lines}) with the versions corrected for
  variations in the intrinsic SEDs ({\em thick lines}), for each bin
  of sSFR.}
\label{fig:intrinsicseds}
\end{figure*}

As noted previously, the procedure we have used to derive the
attenuation curve assumes that the average intrinsic SEDs in the
different bins of Balmer optical depth do not vary substantially from
one another.  Our method of binning the galaxies by sSFR was designed
to mitigate any such intrinsic spectral variations.  However, while
the mean sSFRs as a function of $\tau_{\rm b}$ are similar from
bin-to-bin (Section~\ref{sec:attder} and Figure~\ref{fig:ssfrvstau}),
they are not identical and will therefore introduce some bias in our
estimate of the attenuation curve.

To quantify this bias, we first computed an ``intrinsic'' SED for each
galaxy by: (a) assuming a rising star-formation history with an
e-folding time of $\tau_{\rm r}=5000$\,Myr (this is essentially
equivalent to assuming a constant star-formation history); (b) fixing
the stellar mass of the stellar population to be the value derived
from the SED-fitting (Section~\ref{sec:sedmod}); and (c) varying the
age of the stellar population until the corresponding SFR of the model
is equal to the (Balmer decrement corrected) H$\alpha$-determined one
(Section~\ref{sec:sfrcalc}).  The goal of these steps was to construct
an SED for a stellar population that has the same sSFR as that
computed from SFR(H$\alpha$) and M$^{\ast}$.  Hence, the exact form of
the star-formation history assumed in step (a) above is
inconsequential, as per the discussion in Section~\ref{sec:attder}.

Second, these intrinsic SEDs were normalized so that $f(\lambda =
5500\,{\rm \AA})=1$ and then averaged together for all objects within
each bin of sSFR and $\tau_{\rm b}$ (assuming the same bins used to
derive the attenuation curve).  Figure~\ref{fig:intrinsicseds} shows
these average intrinsic SEDs.  The differences in these intrinsic SEDs
lead to ``correction factors'' $\Delta Q_{\rm int}$ that must be
applied to the $Q_i$ derived above: $\Delta Q_{\rm int} \equiv
1./(\tau_i - \tau_1) \times \ln(f^{\rm int}_i /f^{\rm int}_1)$, where
$f^{\rm int}_i$ is the intrinsic spectrum inferred for the $i^{\rm
  th}$ bin of $\tau_{\rm b}$.  These corrections factors, as a
function of wavelength, are also show in the bottom of the left panel
of Figure~\ref{fig:intrinsicseds}.  

Third, we computed the corrected $Q_i$ as $Q_i^{\rm cor} = Q_i +
\Delta Q_{\rm int}$.  We then followed the same procedure discussed
above to compute $fQ_{\rm eff}^{\rm cor}$: i.e., the $Q_i^{\rm cor}$
in each bin of sSFR were averaged together to produce $Q_{\rm
  eff}^{\rm cor}$ for each sSFR bin, third-order polynomials were fit
to $Q_{\rm eff}^{\rm cor}$, and these functions were multiplied by
the relevant correction factors according to Equation~\ref{eq:multeq}.
Figure~\ref{fig:fqeff} indicates that the $fQ_{\rm eff}^{\rm cor}$ are
essentially identical to the uncorrected versions initially shown in
Figure~\ref{fig:fqeff}.  In effect, correcting for the small
differences in the intrinsic SEDs is mostly compensated for by the
multiplicative factors $f$ which are smaller for the corrected $Q_i$.
The ratios of the uncorrected to corrected $fQ_{\rm eff}$ at $\lambda
= 0.15$ and $2.20$\,$\mu$m are 1.083 and 0.998, respectively, for sSFR
bin 1; and 0.992 and 1.056, respectively, for sSFR bin 2.  Given the
less than $\approx 8\%$ differences implied for the total attenuation
curve, we assume the uncorrected version expressed in
Equation~\ref{eq:kstars}.

Finally, we note that systematic offsets in the derived stellar masses
or SFR(H$\alpha$) resulting from the assumption of different
attenuation curves in the SED fitting will simply shift all the
galaxies up or down in terms of sSFR, but the {\em ordering} of the
galaxies will for the most part be preserved.\footnote{The ordering of
  galaxies will be preserved if the same attenuation curve applies to
  all galaxies.  This is the case for galaxies in our sample given the
  similarity in shape between $Q_2$ and $Q_3$ in a given sSFR bin
  (Figure~\ref{fig:selq}), and the similarly in shape of $fQ_{\rm
    eff}$ between the two sSFR bins (Figure~\ref{fig:fqeff}).}  For
this reason, the ratios of the intrinsic SEDs shown in
Figure~\ref{fig:intrinsicseds} will be unaffected by assuming a
different attenuation curve in the derivation of SFR(H$\alpha$) and
$M^{\ast}$.  Detailed comparisons of the SED parameters and SFRs
derived with different assumptions of the attenuation curve are
presented in Section~\ref{sec:discussion}.

\subsubsection{Comparison to Other Common Curves}

The average attenuation curve found for the MOSDEF galaxies is
compared to several local line-of-sight extinction curves and the
\citet{calzetti00} attenuation curve in Figure~\ref{fig:comparek}.
The SMC, LMC, and Milky Way curves, and their respective
normalizations, are taken from \citet{gordon03}.  A common feature of
these curves is that they all asymptote to zero as the dust absorption
cross-section becomes negligible at long wavelengths, with a
corresponding rapid rise in the attenuation curve toward shorter (UV)
wavelengths.  The one distinguishing feature of the Milky Way and LMC
extinction curves, namely the $2175$\,\AA\, ``bump'' \citep{stecher65,
  fitzpatrick86}, is absent from the attenuation curve found for local
UV starburst galaxies \citep{calzetti00}, but its presence to varying
degrees has been noted in higher-redshift galaxies (e.g.,
\citealt{noll09, buat11, buat12, kriek13, scoville15}).

\begin{figure}
\epsscale{1.}
\plotone{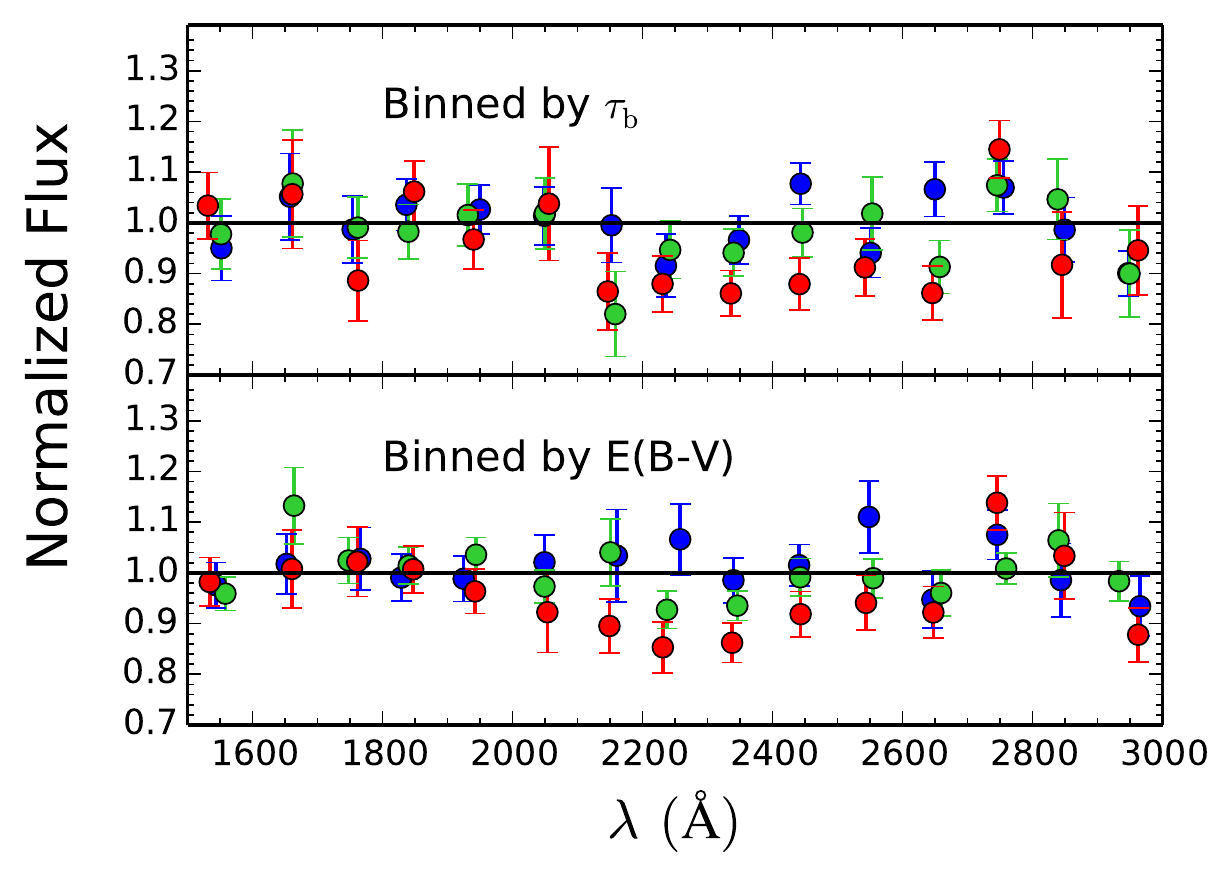}
\caption{Normalized average photometry as a function of rest-frame
  wavelength for galaxies in bins of Balmer optical depth ({\em top})
  and color excess of the stellar continuum ({\em bottom}).  The
  average photometry of galaxies is computed in bins of rest-frame
  wavelength of $\Delta\lambda=100$\,\AA, and the average photometry
  is normalized using a linear fit to the points over wavelength
  ranges $\lambda = 1500-1800$ and $2600-3000$\,\AA.}
\label{fig:bump}
\end{figure}

To ascertain the existence of this excess absorption, we examined the
average photometry of galaxies in bins of $\tau_{\rm b}$ and $E(B-V)$,
where the latter were determined from the SED modeling discussed in
Section~\ref{sec:sedmod}.  To improve the rest-frame wavelength
sampling of the average photometry, we include all galaxies
irrespective of their sSFRs.  We then calculated the average, and
error in the average, flux densities in rest-frame wavelength bins of
$100$\,\AA (e.g., similar to the method used above to compare the
average photometry of galaxies in our sample to the functional fit to
the photometry; Figure~\ref{fig:aveseds}).  A linear function was fit
to the log of the average flux density as a function of
$\log(\lambda)$ in two wavelength ranges bracketing the $2175$\,\AA\,
feature: $\lambda = 1500-1800$ and $2600-3000$\,\AA.  The errors in
the mean flux densities were used to weight the points in the fitting.
The average flux densities throughout the wavelength range from $1500$
to $3000$\,\AA\, were then divided by this linear function to produce
the normalized points shown in Figure~\ref{fig:bump}.

There is on average a lower flux at wavelengths corresponding to
roughly where the excess absorption would be expected.\footnote{As the
  wavelength windows used to compute the UV slope largely exclude the
  region where we observe a deficit in the average normalized flux,
  the effect of imposing a linear fit to the flux as a function of
  wavelength will result in a minimal bias of $\Delta \beta \la 0.1$.
  This small bias does not affect our conclusions regarding the
  contribution of A stars to the near-UV continuum
  (Section~\ref{sec:astars}), though of course fitting the slope
  without the photometry that may be affected by the $2175$\,\AA\,
  absorption will result in larger random uncertainties.}  While the
depth of the absorption appears to be insensitive to $\tau_{\rm b}$,
there is some marginal significance for a trend between the absorption
depth and the color excess of the stellar continuum.  A Spearman
correlation test implies that that the null hypothesis of no
correlation between the depth of the $2175$\,\AA\, and the color
excess of the stellar continuum can be ruled out at the $\simeq
3$\,$\sigma$ level.  We caution that the excess absorption will be
smoothed by the finite wavelength sampling of the average photometry
as well as the fact that the average photometry is computed from
broad- and intermediate-band magnitudes.

For comparison, \citet{buat12} deduce the shape of the dust
attenuation curve for a sample of 751 galaxies at $0.95<z<2.2$, and
find evidence for excess absorption at 2175\,\AA\, for roughly $20\%$
of their sample, almost all ($90\%$) of which lie at $z<1.5$.  They
also find that the rate of a secure detection of bump increases by a
factor of $\approx 2$ when considering the dustier galaxies (i.e.,
those directly detected at mid- and far-IR wavelengths).  In another
recent study, \citet{kriek13} examine a sample of somewhat
lower-redshift ($0.5<z<2.0$) galaxies and find that the strength of
the $2175$\,\AA\, absorption correlates with the ``steepness'' (i.e.,
the wavelength dependence) of the attenuation curve.  However, we note
that the galaxies in their sample have $W_{\rm H\alpha}<140$\,\AA,
values that are substantially lower (corresponding to lower sSFRs)
than those considered here ($W_{\rm H\alpha}\ga 100$\,\AA).  More
definitive results on the extent to which the $2175$\,\AA\, absorption
may be present in the galaxies targeted with MOSDEF, as well as the
characteristics (e.g., slope of the attenuation curve, sSFR, redshift)
upon which its depth may depend, will come from the larger (completed)
survey dataset.

Regardless, our analysis shows that the attenuation curve for the
MOSDEF galaxies exhibits a slope that is identical to that of the
\citet{calzetti00} attenuation curve for $\lambda < 0.6$\,$\mu$m, but
with a lower normalization ($R_V$).  For $\lambda > 0.6$\,$\mu$m, the
MOSDEF curve is similar to that of the SMC curve, and more generally
the two are essentially identical for $\lambda > 0.25$\,$\mu$m.
Recent studies have suggested that some populations of high-redshift
galaxies have reddening consistent with the SMC (e.g.,
\citealt{reddy06a, siana08, siana09, reddy10, shim11, oesch13}).
Here, we have shown that through direct measurements of the
attenuation curve, an SMC-like behavior at longer wavelengths
($\lambda > 2500$\,\AA) may be more generally applicable at high
redshift.  A comparison between physical models of the distribution of
dust and stars in galaxies and the attenuation curve will be presented
elsewhere, but we note that the shape of the attenuation curve may
imply a larger covering fraction of dust in the rest-optical, or a
different dust grain size distribution, than would be indicated by the
\citet{calzetti00} curve.  The implications for these differences in
the shapes of the attenuation curves are discussed in the next
section.

\section{Discussion and Implications}
\label{sec:discussion}

\subsection{Reassessment of the Stellar Populations}
\label{sec:sedcompare}

\begin{figure*}[tbh]
\epsscale{1.}
\plotone{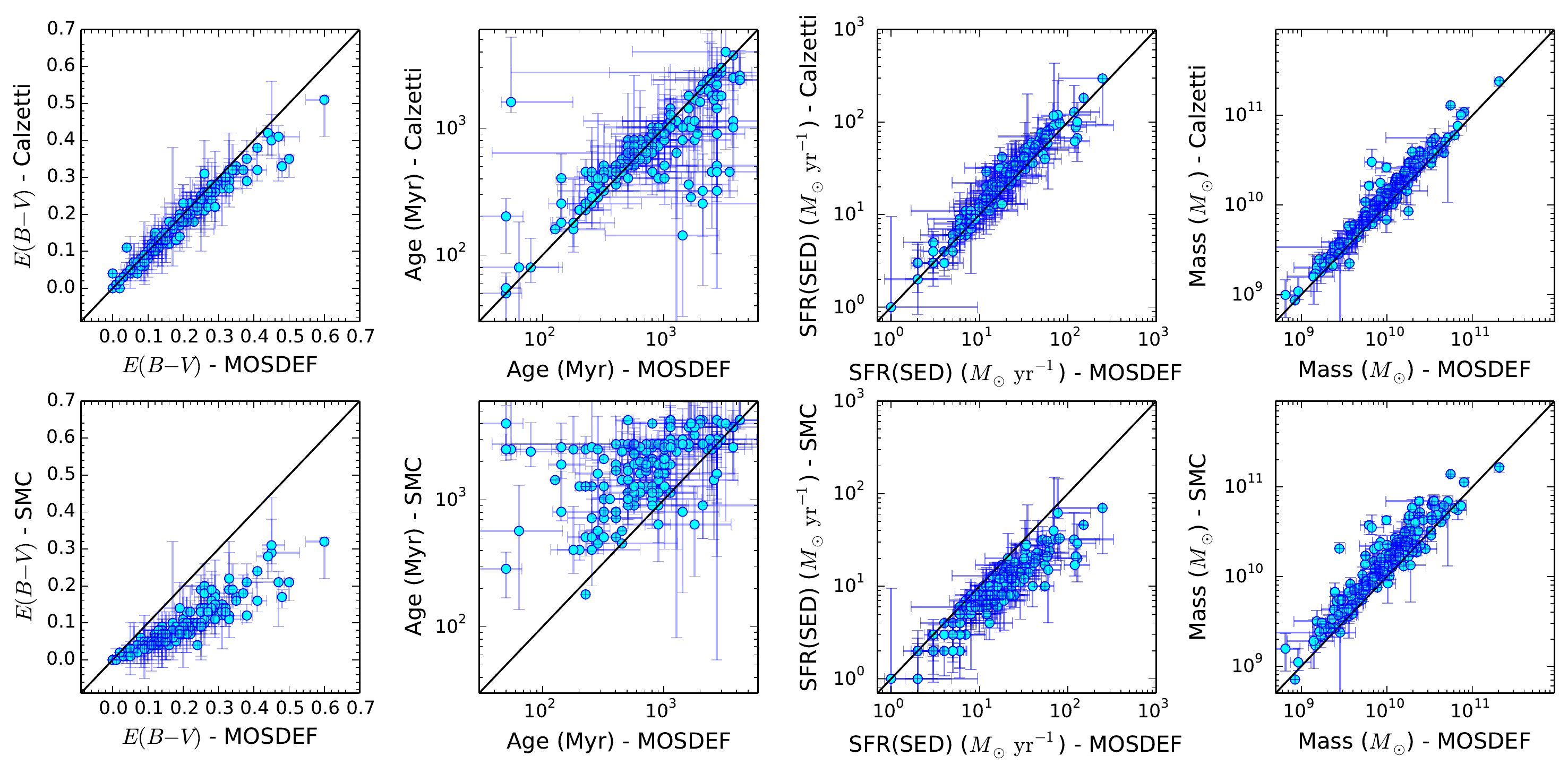}
\caption{Comparison of the color excesses, ages, SFRs, and stellar
  masses returned from SED fitting assuming our new attenuation curve
  relative to those obtained with the \citet{calzetti00} and SMC
  attenuation curves.  The solid lines denote a one-to-one agreement
  between the plotted quantities.}
\label{fig:sedcompare}
\end{figure*}

Given the apparent differences in the attenuation curve derived in
this work relative to the curve typically assumed for high-redshift
galaxies (i.e., \citealt{calzetti00}), we adopted our new calculation
of the attenuation curve to refit the stellar populations of galaxies
in our sample, in the same way as described in
Section~\ref{sec:sedmod}.  The comparison of several stellar
population parameters using our new curve versus those obtained with
the \citet{calzetti00} and SMC curves is shown in
Figure~\ref{fig:sedcompare}.  The mean color excess of the stellar
continuum assuming the MOSDEF attenuation curve is $\langle \Delta
E(B-V)\rangle \approx 0.02$ redder than that obtained with the
\citet{calzetti00} relation ($E(B-V)\approx 0.19$ vs. $0.17$).
Combining this average difference in the color excesses with the
$\Delta k\approx 1.63$ difference in the two attenuation curves at
$\lambda = 1600$\,\AA\, we infer dust obscuration factors (required to
de-redden the observed UV SFRs) that are $10^{0.09}\approx 1.2$ times
smaller than those inferred from the \citet{calzetti00} relation.  As
a result, the average SED-inferred SFRs obtained with the MOSDEF
attenuation curve are $\approx 20\%$ smaller than those obtained with
a \citet{calzetti00} curve.  This systematic difference in the SFRs is
typically smaller than the random uncertainties, where the latter were
computed using Monte Carlo realizations of the observed photometry and
refitting the model SEDs to the perturbed photometry;
Section~\ref{sec:sedmod}).  Similarly, the systematic offset seen here
is well within the typical factor of $\approx 2$ uncertainty in
independent measurements of the SFRs based on combining UV and IR data
for galaxies with similar SFRs and stellar masses, and at the same
redshifts, as those considered here \citep{reddy06a, daddi07a,
  reddy10, reddy12a}.  These results apply on average, as it is clear
from Figure~\ref{fig:sedcompare} that the difference in $E(B-V)$
implied by the two attenuation curves is a function of the ``redness''
of the object.  For the bluest galaxies, we find $\Delta E(B-V)\approx
0$ (these are galaxies for which little reddening is required to
reproduce the UV colors).  For the reddest galaxies (e.g., those with
$E(B-V)>0.3$ as computed from the MOSDEF attenuation curve), we find
$\Delta E(B-V) \approx 0.08$.

\begin{figure}[tbh]
\epsscale{1.}
\plotone{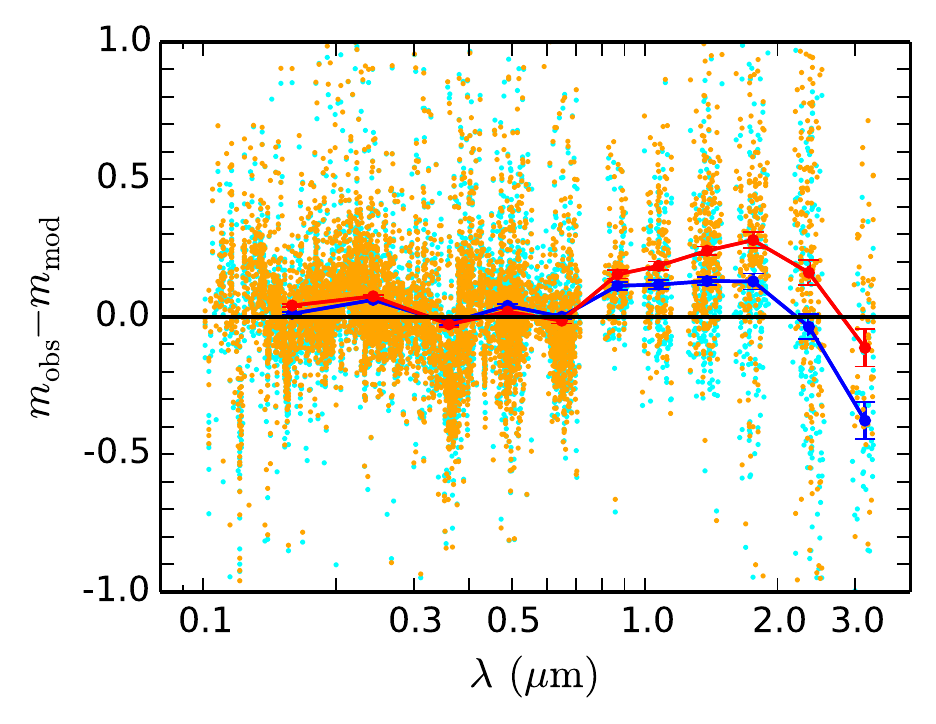}
\caption{Residuals (i.e., difference in the observed and model
  photometry) for galaxies in our sample assuming a rising
  star-formation history and the MOSDEF ({\em cyan}) and
  \citet{calzetti00} ({\em orange}) attenuation curves.  The mean
  residuals and the errors in these means are indicated by the {\em
    blue} and {\em red} symbols assuming the MOSDEF and
  \citet{calzetti00} curves, respectively.  For rest-frame wavelengths
  between $\approx 1$ and $2.5$\,$\mu$m, The MOSDEF attenuation curve
  results in a statistically significant improved fit to the observed
  photometry, on average, relative to that obtained assuming the
  \citet{calzetti00} curve.}
\label{fig:residuals}
\end{figure}

Assuming the MOSDEF attenuation curve results in galaxies that are on
average $\approx 300$\,Myr older relative to the ages derived using
the \citet{calzetti00} curve, but we view these differences to be
negligible given the large uncertainties in the ages.\footnote{A
  subset of objects have ``best-fit'' ages that are a factor of $\ga
  2$ older with the MOSDEF attenuation curve relative to those
  obtained with the \citet{calzetti00} curve.  These are objects where
  the MOSDEF and \citet{calzetti00} ``best-fit'' SEDs are essentially
  identical in the rest-frame optical, but where the former are
  generally fainter in the rest-frame near-IR.  As such, the implied
  Balmer/4000\,\AA\, breaks are stronger for the MOSDEF ``best-fit''
  SEDs resulting in older ages.}  Note that the ``uncertainties''
quoted here do not account for other systematic effects, including
differences in the assumed star-formation histories of galaxies.
Finally, we note that assuming the MOSDEF attenuation curve yields
better fits to the longer wavelength (i.e., rest-frame $\simeq
1.0-2.5$\,$\mu$m) photometry than the \citet{calzetti00} curve for
rising star-formation histories (Figure~\ref{fig:residuals}).  As
such, the stellar masses are $\approx 25\%$ lower when we adopt the
MOSDEF attenuation curve in lieu of the \citet{calzetti00} curve.
This implies an $\approx 0.1$\,dex difference in $\log M^\ast$ that is
of the same order as the uncertainty in stellar mass implied by the
photometric errors.  Similarly, in comparison to the results derived
with an SMC attenuation curve, those assuming the MOSDEF curve imply
significantly redder E(B-V) ($\Delta E(B-V)\approx 0.10$), younger
ages (by a factor of $\approx 2$), higher SFRs (by a factor of
$\approx 2$), and lower stellar masses ($\Delta \log (M^\ast/M_\odot)
\approx 0.16$\,dex).  The differences stem from the fact that the SMC
curve has a steeper wavelength dependence in the UV than either the
MOSDEF or \citet{calzetti00} attenuation curves, and thus a lower
amount of reddening is required to fit a given UV color.  As a result,
the SFRs are lower with the SMC curve.  Further, the near-IR-optical
color is attributed more to an older stellar population than dust
reddening, resulting in older ages and larger stellar masses.

\begin{figure}[tbh]
\epsscale{1.}
\plotone{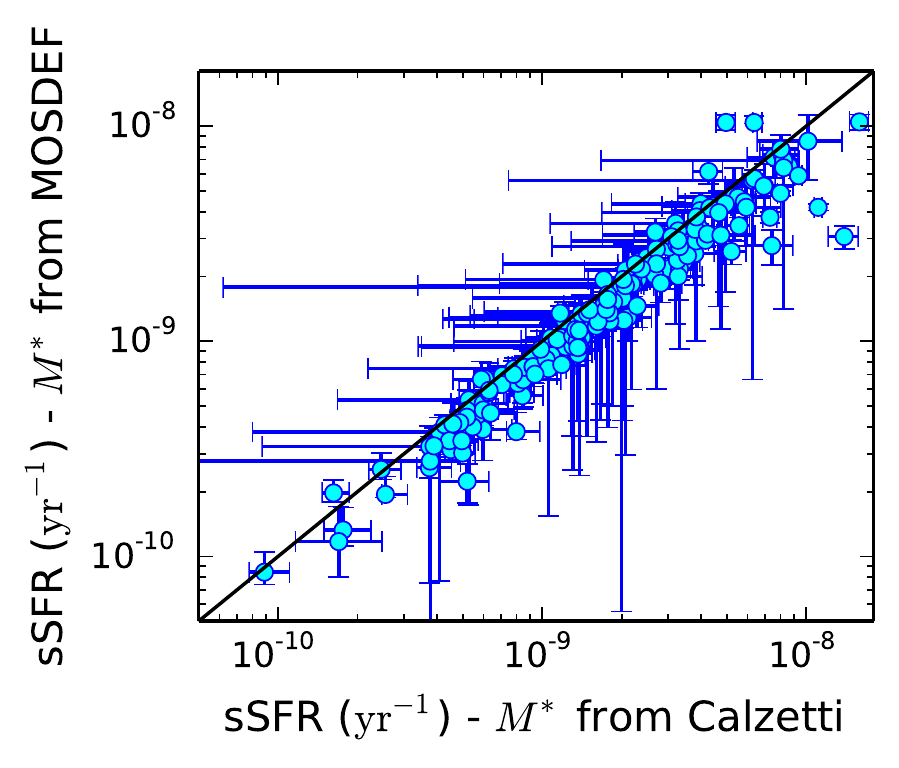}
\caption{Specific SFRs calculated assuming H$\alpha$-based SFRs, and
  stellar masses from SED fitting.  SFR(H$\alpha$) assumes
the \citet{cardelli89} extinction curve.  The stellar masses
assume the \citet{calzetti00} (ordinate) and the 
MOSDEF (abscissa) attenuation curves.}
\label{fig:ssfrcompare}
\end{figure}

Another point of consideration is how the sSFRs are affected by
different assumptions of the attenuation curve.  Recall that the sSFRs
were calculated by combining H$\alpha$-based SFRs, which were
determined independently of SED fitting, with the SED-based stellar
masses (Section~\ref{sec:sedmod}).  The H$\alpha$-based SFRs were
derived assuming the \citet{cardelli89} extinction curve
(Section~\ref{sec:sfrcalc}).  The SED-based stellar masses were
computed assuming the \citet{calzetti00} curve.  We will return to a
more detailed discussion of the implications of our results for SFR
determinations (Section~\ref{sec:sfrimplications}), but for the
remainder of our analysis, we adopt the sSFRs computed assuming the
stellar masses obtained with the MOSDEF attenuation curve, and
SFR(H$\alpha$) obtained with the \citet{cardelli89} extinction curve.
As shown in Figure~\ref{fig:ssfrcompare}, the new sSFRs computed in
this way are quite similar to those computed when we assume the
\citet{calzetti00} attenuation curve in deriving the stellar masses,
with the former being $\Delta\log({\rm sSFR/yr^{-1}}) \simeq
0.09$\,dex smaller than the latter.  Because there is only a
systematic offset between the two sSFR determinations, and the
relative ordering of the galaxies is preserved, the sSFR bins used to
calculate the attenuation curve (Section~\ref{sec:attder}) will
include essentially the same set of galaxies regardless of which
attenuation curve is assumed (and, in any case, the results of
Section~\ref{sec:attder} suggest that because the attenuation curves
computed from the two sSFR bins are very similar, we could have simply
ignored any binning with sSFR in computing the curve).

\subsection{Measurements of UV Slopes}
\label{sec:betamods}

As noted in Section~\ref{sec:astars}, $\beta_{\rm
  phot}(1260-1750\,{\rm \AA})$ is not substantially bluer than
$\beta_{\rm phot}(1260-2600\,{\rm \AA})$, suggesting that A stars do
not contribute significantly to the UV continuum for the galaxies in
our sample.  It is worthwhile to consider this point further, because
an intrinsically blue spectral slope in the range $1260-1750$\,\AA\,
can appear redder than the slope measured over the wider baseline in
wavelength ($1260-2600$\,\AA) if the attenuation curve is rising
steeply in the UV.  Therefore, we must rule out the case where the
attenuation curve masks the contribution from A stars to the near-UV
continuum.  Our new derivation of the attenuation curve implies that
the difference between $\beta(1260-1750\,{\rm \AA})$ and
$\beta(1260-2600\,{\rm \AA})$ for galaxies where the latter is $<0.5$
(a limit that encompasses almost all of the galaxies in our sample),
is $\Delta \beta < 0.2$ for $E(B-V) \la 0.4$.  This difference in UV
slope is small compared to both the typical measurement errors in the
UV slopes and the dispersion in $\beta(1260-1750\,{\rm \AA})$ at a
given $\beta(1260-2600\,{\rm \AA})$
(Figure~\ref{fig:betabluecompare}).  Thus, the reddening implied by
the MOSDEF attenuation curve is insufficient to make an intrinsically
blue $\beta(1260-1750\,{\rm \AA})$ appear substantially redder than
$\beta(1260-2600\,{\rm \AA})$, and our conclusion regarding the
contribution of A stars to the UV continuum still holds.

The test just discussed allows us to calculate the useful relationship
between $\beta$ and the color excess of the stellar continuum assuming the
MOSDEF attenuation curve and an underlying solar metallicity stellar
population with constant star formation for at least $100$\,Myr (i.e.,
with an intrinsic spectral slope of $\beta_{\rm int}\approx -2.44$):
$\beta = -2.44 + 4.54\times{\rm E(B-V)}$.  Additionally,
Equations~\ref{eq:kdef} and \ref{eq:kstars} imply that the 
attenuation in magnitudes at $1600$\,\AA\, is related to the color
excess by $A_{\rm 1600\,\mbox{\AA}} = 8.34\times {\rm E(B-V)}$.  Thus,
combining the previous two equations, we find
\begin{equation}
A_{\rm 1600\,\mbox{\AA}} = 1.84\beta + 4.48,
\label{eq:magext}
\end{equation} 
again for an intrinsic spectral slope $\beta_{\rm int} \simeq -2.44$.
For comparison, the relation from \citet{meurer99} is $A_{\rm
  1600\,\mbox{\AA}} = 1.99\beta + 4.43$, or $A_{\rm 1600\,\mbox{\AA}}
= 1.99\beta + 4.86$ when shifted to account for the difference in the
intrinsic UV slope ($\beta_0$) between \citet{meurer99} and the
present study (i.e., $\beta_0=-2.23$ vs $-2.44$).  Consequently, for
the bulk of the galaxies in our sample with $\beta<-0.5$, the implied
difference in the magnitudes of extinction at $1600$\,\AA, when
assuming the MOSDEF and \citet{meurer99} attenuation curves, is
$\Delta A_{\rm 1600\,\mbox{\AA}} \la 0.31$.  More specifically, the
MOSDEF attenuation curve implies dust obscuration factors,
$10${\textrm \^{[0.4$\times A$(1600\,\mbox{\AA})]}}, that are a factor
of $\la 1.32$ times lower than those implied by the \citet{meurer99}
(or \citealt{calzetti00}) attenuation curve, with an average
correction that results in SFRs that are $\approx 20\%$ lower for the
MOSDEF curve, as noted above.  Equation~\ref{eq:magext} can be used to
directly determine the attenuation of the UV continuum for galaxies of
a given (measured) UV slope.  However, as we discuss in
Section~\ref{sec:sfrimplications}, additional factors may be required
to recover total SFRs from UV-based ones.

\subsection{Comparison of the Color Excesses and Total 
Attenuation of the Stellar Continuum and Ionized Gas}
\label{sec:gasvsstars}

\begin{figure*}
\epsscale{1.}
\plotone{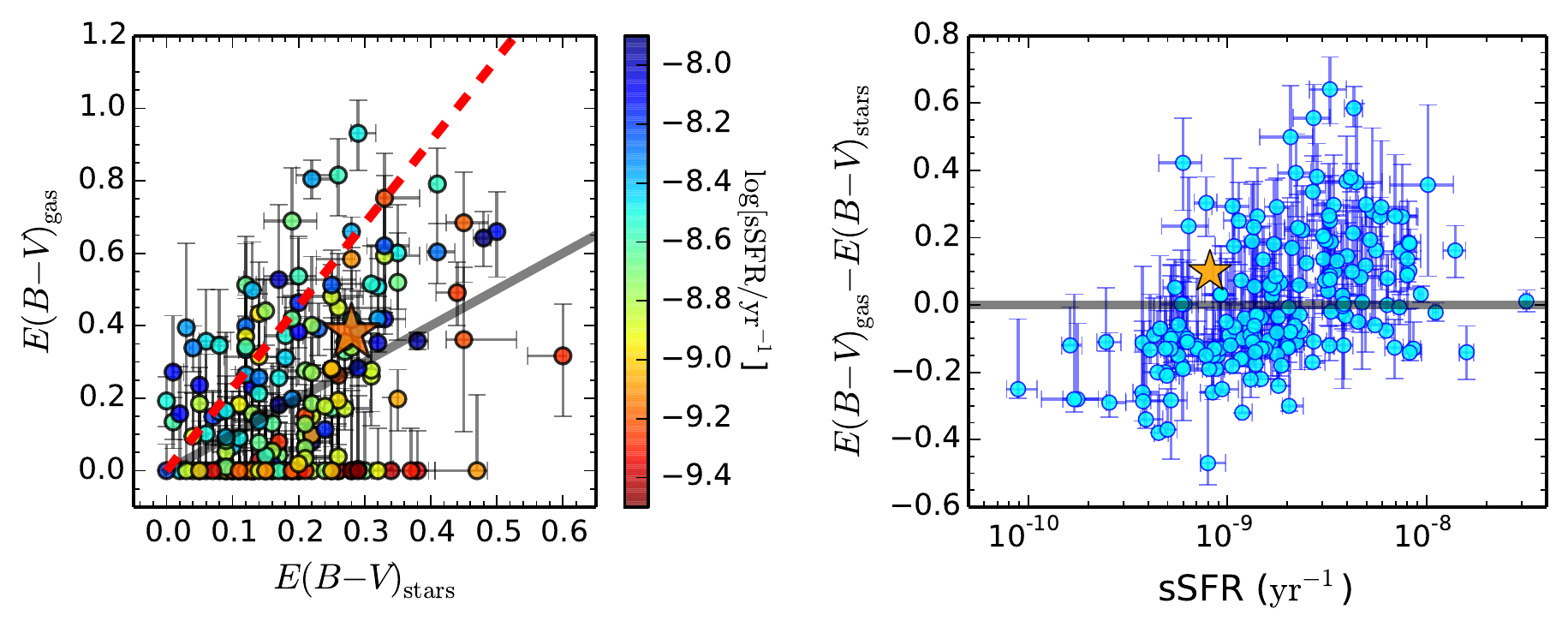}
\caption{{\em Left:} Comparison of the color excesses derived for the
  stellar continuum and the ionized gas.  $E(B-V)_{\rm gas}$ is
  computed assuming the \citet{cardelli89} extinction curve, and
  $E(B-V)_{\rm stars}$ is the value returned from the SED fitting when
  we assume the MOSDEF attenuation curve.  The solid line denotes
  identical color excesses derived for the stellar and gas components,
  and the dashed line indicates the relation $E(B-V)_{\rm gas} =
  E(B-V)_{\rm stars}/0.44$ from \citet{calzetti00}.  Points are
  differentiated according to the sSFR.  The large star denotes the
  average values for the H$\beta$-undetected galaxies.  {\em Right:}
  Difference between gas and continuum color excesses as a function of
  sSFR, where the solid line indicates no difference between the color
  excess of the nebular regions and the stellar continuum.  The large
  star denotes the average values for the H$\beta$-undetected
  galaxies.}
\label{fig:ebmvcompare}
\end{figure*}

There has been a considerable effort to investigate the relationship
beteween the attenuation of the stellar continuum and ionized gas in
high-redshift galaxies, particularly in light of local observations of
star-forming galaxies (e.g., \citealt{fanelli88, calzetti94,
  mashesse99, moustakas06, kreckel13}) that suggest a higher reddening
of the ionized gas than of the stellar continuum : e.g., $E(B-V)_{\rm
  stars} = 0.44\times E(B-V)_{\rm gas}$ \citep{calzetti00}.  Note that
this often-quoted relationship describes the color excess of the
stellar continuum and the nebular line-emitting regions, not the total
attenuation.  More precisely, this relation is derived under the
assumption that the starburst attenuation curve applies to the stellar
continuum, and a Galactic extinction curve \citep{cardelli89} to the
ionized gas \citep{calzetti01}.  However, most investigations of the
stellar and nebular reddening in high-redshift galaxies have assumed
that the same attenuation curve applies for both the gas and the
stars, and it is unclear if this is an appropriate assumption in the
context of high-redshift galaxies.

From the standpoint of local galaxies, the difference in attenuation
curves can be understood in the framework where the recombination
emission arises from spatially compact HII regions that are subject to
a high covering fraction of dust and where a line-of-sight extinction
curve (e.g., \citealt{cardelli89} or SMC) is more appropriate, versus
the more spatially extended stellar continuum that has a lower dust
covering fraction and is subject to a ``greyer'' attenuation curve
(e.g., \citealt{calzetti00}).  This configuration is similar to the
two component dust model of \citet{charlot00} where the HII regions
are attenuated by a component of dust in undissipated parent birth
clouds, with a second diffuse component of dust in the ISM.

Studies of UV-selected galaxies at $z\sim 2$ have shown a general
agreement between UV and H$\alpha$-based SFRs when applying the same
color excess and assuming the same attenuation curve applies for both
the stars and gas (e.g., \citealt{erb06c, reddy10}; see also
\citealt{cowie08} for results at $z<0.5$).  On the other hand, studies
that have targeted on average more highly star-forming or more massive
galaxies, and/or photometrically-selected galaxies, indicate a higher
attenuation of the gas versus the stellar continuum, again with most
studies assuming that the same attenuation curve applies for both the
stars and gas (e.g., \citealt{forster09, yoshikawa10, wuyts11,
  kashino13, wuyts13, price14}).

\subsubsection{Comparison of Color Excesses}

A plausible scenario that can resolve the current observations is one
in which the difference in the continuum and nebular attenuation is a
function of SFR, mass, or sSFR (e.g., \citealt{reddy10, price14}).  As
remarked in Section~\ref{sec:nebvsstel}, the selective attenuation
curve derived in this study implies that {\em on average} the ionized
gas suffers greater extinction than the continuum (at the wavelengths
of H$\alpha$ and H$\beta$).  We scrutinized this result by comparing
the color excesses of the stellar continuum derived from SED fitting,
with those of the ionized gas, computed as:
\begin{eqnarray}
E(B-V)_{\rm gas} & = & \frac{2.5}{k({\rm H}\beta)-k({\rm H}\alpha)}
\log_{10}\left(\frac{{\rm H}\alpha/{\rm H}\beta}{2.86}\right) \nonumber \\
& = & \frac{1.086}{k({\rm H}\beta)-k({\rm H}\alpha)}\tau_{\rm b} 
\simeq 0.950\tau_{\rm b}.
\label{eq:ebmvgas}
\end{eqnarray}
In the last step, we have assumed that the \citet{cardelli89}
extinction curve applies to the HII regions.  Adopting the
\citet{calzetti00} and MOSDEF attenuation curves results in
$E(B-V)_{\rm gas}$ values that are a factor of $0.90\times$ and
$1.01\times$ those assuming the \citet{cardelli89} curve,
respectively.  There is a large scatter in the relationship between
the gas and continuum color excesses, with $\approx 50\%$ of the
objects having equal gas and continuum color excesses within the
($1$\,$\sigma$) uncertainties, and only $7\%$ consistent with having
$E(B-V)_{\rm gas}>E(B-V)_{\rm stars}/0.44$.  As could have been
anticipated from Figures~\ref{fig:beta_vs_taub} and \ref{fig:linfit},
the degree to which $E(B-V)_{\rm gas}$ diverges from $E(B-V)_{\rm
  stars}$ is dependent upon the sSFR of the galaxy, with the largest
differences for galaxies with the largest sSFRs
(Figure~\ref{fig:ebmvcompare}).\footnote{As we show in
  Section~\ref{sec:prescription}, a {\em different} relation is found
  between $E(B-V)_{\rm gas}-E(B-V)_{\rm stars}$ and sSFR(SED), i.e.,
  the sSFR computed when we use {\em SED-based} SFRs.}

\begin{figure*}[htb]
\epsscale{1.}
\plotone{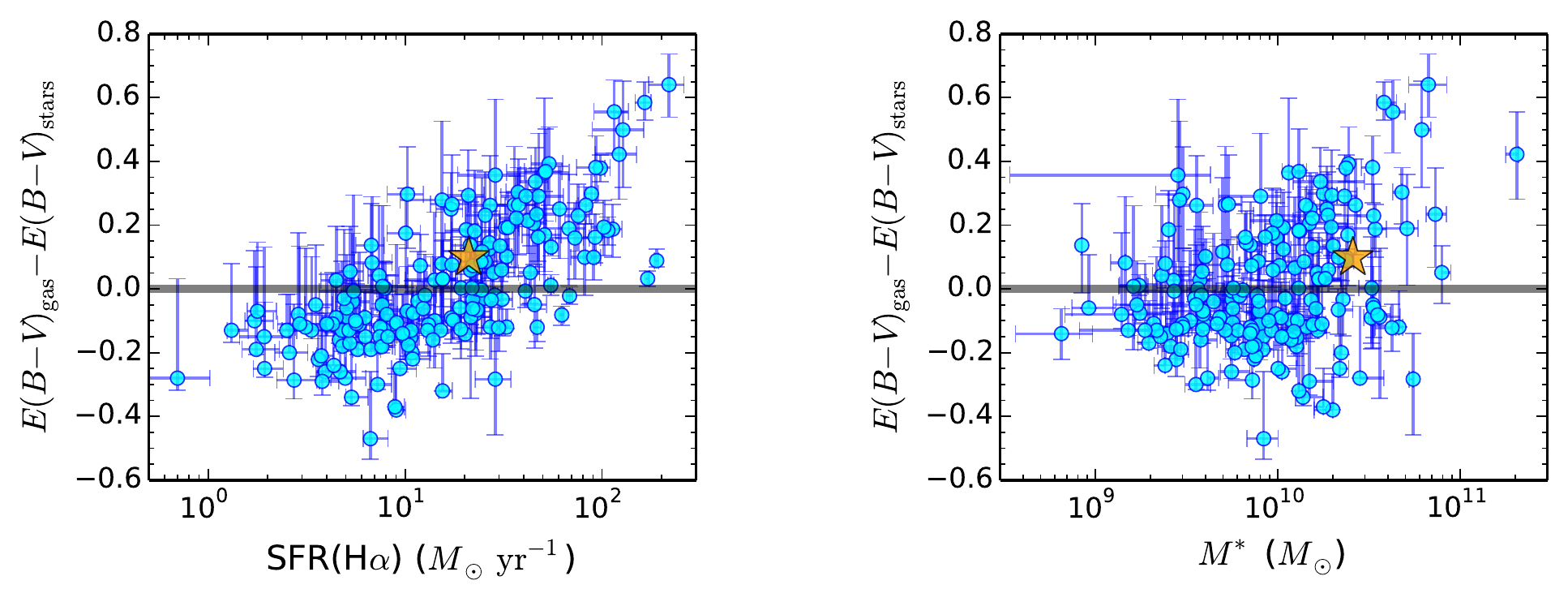}
\caption{Difference between gas and continuum color excesses as a
  function of H$\alpha$ SFR ({\em left}) and stellar mass ({\em
    right}), where the solid lines indicate no difference between the
  color excess of the nebular regions and the stellar continuum, and
  the large stars indicate the values for H$\beta$-undetected
  galaxies.  SFR(H$\alpha$) is computed assuming the
  \citet{cardelli89} extinction curve.}
\label{fig:ebmvcompare_sfr}
\end{figure*}

We should consider, however, whether selection biases could result in
correlation between the difference in the gas and nebular color
excesses and the sSFR.  In particular, if galaxies with H$\alpha$
fluxes close to our detection limit of $\simeq 5\times
10^{-18}$\,erg\,s$^{-1}$\,cm$^{-2}$ are dusty, then their H$\beta$
lines may be undetected with a greater frequency than galaxies of the
same dust content but with bright H$\alpha$ lines.  As a result, we
may be insensitive to galaxies with low SFRs and high nebular color
excesses.  However, as seen in Figure~\ref{fig:bd_vs_lha}, the
H$\beta$-undetected galaxies in our sample have limits that span the
full range of observed $L_{\rm H\alpha}$.  Further, if we compute the
average SFR(H$\alpha$), sSFR, $E(B-V)_{\rm gas}$, and $E(B-V)_{\rm
  stars}$ for the H$\beta$-undetected galaxies, as shown in
Figures~\ref{fig:ebmvcompare} and \ref{fig:ebmvcompare_sfr} (where the
averages for the SFR and $E(B-V)_{\rm gas}$ are determined from the
composite stack of their spectra; Section~\ref{sec:sample}), the
aforementioned quantities suggest that there is not a large population
of dusty galaxies with low SFRs and low sSFRs.  Aside from this direct
observation, it has been shown in several studies that the mean
obscuration of high-redshift galaxies increases with SFR (e.g.,
\citealt{adelberger00,reddy06a,buat07, burgarella09, reddy08, buat09,
  reddy10,sobral12, dominguez13}; see also discussion below),
suggesting that among galaxies with low H$\alpha$ fluxes, those that
are intrinsically less dusty vastly outnumber those with high dust
content.  The stacking of mid- to far-IR data for the MOSDEF galaxies
will help to clarify these arguments.  There is no similar bias in
terms of $E(B-V)_{\rm stars}$, as the spectroscopically confirmed
MOSDEF galaxies have rest-UV colors similar to those of galaxies in
the parent sample \citep{kriek14}.

Formally, a Kendall's $\tau$ test indicates a probability of $\la
1.8\times 10^{-7}$ that the difference between gas and stellar color
excesses is {\em uncorrelated} with sSFR.  The correlation with
H$\alpha$-based SFR is even stronger, with $\la 10^{-9}$ probability
that $E(B-V)_{\rm gas}-E(B-V)_{\rm stars}$ is uncorrelated with SFR
(Figure~\ref{fig:ebmvcompare_sfr}).  Similarly, there is a $\la
10^{-4}$ probability that the difference in color excesses is
uncorrelated with stellar mass.  Thus, the conflicting results on the
relationship between the color excesses of the HII regions and stellar
continuum in high-redshift galaxies can be reconciled if selection
over different ranges of SFR and/or stellar mass leads to the
inclusion of more or fewer galaxies with the largest discrepancies
between nebular and continuum color excess.  Specifically, galaxies
with SFRs$\ga 20$\,$M_\odot$\,yr$^{-1}$ exhibit nebular color excesses
that are substantially redder than the continuum color excesses.

In contrast, over the range of SFRs characteristic of UV-selected
galaxies at $z\sim 2$ ($\la 20$\,$M_\odot$\,yr$^{-1}$ for a Chabrier
IMF; e.g., \citealt{reddy12b}), the nebular color excesses are
essentially equivalent to or bluer than those of the stellar
continuum.  Of the 51 objects that have $E(B-V)_{\rm gas}$ that is
more than $1$\,$\sigma$ bluer than $E(B-V)_{\rm stars}$, 35 (or
$\approx 69\%$) formally have $E(B-V)_{\rm gas} = 0$.  These galaxies
are unlikely to have HII regions that are completely dust-free given
the reasoning presented in Section~\ref{sec:betavstauscatter}, and
their $E(B-V)_{\rm stars}$ span the full range observed for galaxies
in our sample in general.  An analysis of the spatially-resolved
colors of these galaxies should lend more insight into the question of
why they exhibit $E(B-V)_{\rm gas}$ that are bluer than the color
excess of the stellar continuum.  In summary, we find that the
difference in the color excesses of the stellar continuum and ionized
gas in high-redshift galaxies is a strong function of SFR: above an
SFR of $\approx 20$\,$M_\odot$\,yr$^{-1}$, the nebular lines become
increasingly reddened relative to the color excess of the stellar
continuum.

\subsubsection{Comparison of the Total Attenuation of the Ionized Gas
and Stellar Continuum}

\begin{figure}[htb]
\epsscale{1.}
\plotone{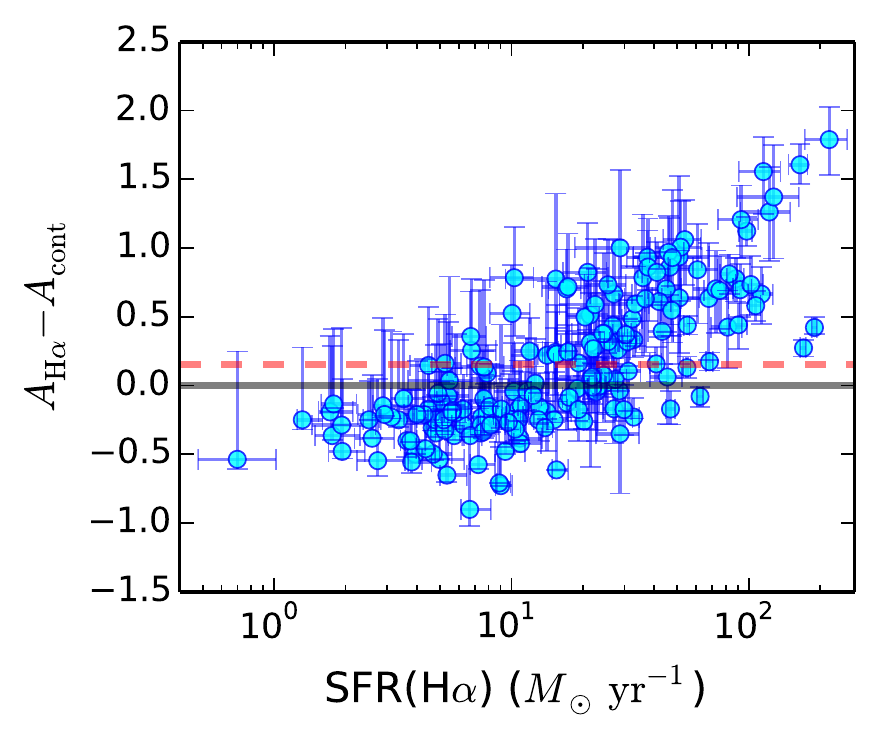}
\caption{Difference between the magnitudes of attenuation of
  recombination line photons and continuum photons at $6565$\,\AA, as
  a function of SFR.  $A_{\rm H\alpha}$ and $A_{\rm cont}$ assume the
  \citet{cardelli89} Galactic extinction curve and the MOSDEF
  attenuation curve, respectively.  The solid and dashed lines denote
  equal attenuation and the mean difference ($\langle A_{\rm H\alpha}
  - A_{\rm cont}\rangle \simeq 0.15$) in attenuation of line and
  continuum photons, respectively.}
\label{fig:compatt}
\end{figure}

So far, our discussion has focused on a comparison of the {\em color
  excesses} of the ionized gas and the stellar continuum in
high-redshift galaxies.  Recall that the color excess of the ionized
gas was computed assuming the \citet{cardelli89} extinction curve.
The color excess of the stellar continuum was derived assuming the
MOSDEF attenuation curve.  These respective curves were used to also
compute the total attenuation in magnitudes of the line and continuum
emission at 6565\,\AA:
\begin{eqnarray}
A_{\rm H\alpha} & = & k_{\rm Cardelli}(6565\,{\rm \AA}) \times E(B-V)_{\rm gas} \nonumber \\
& = & 2.52\times E(B-V)_{\rm gas}; \nonumber \\
A_{\rm cont} & = & k_{\rm MOSDEF}(6565\,{\rm \AA}) \times E(B-V)_{\rm stars} \nonumber \\
& = & 1.92\times E(B-V)_{\rm stars}.
\end{eqnarray}
The comparison of $A_{\rm H\alpha}$ and $A_{\rm cont}$
(Figure~\ref{fig:compatt}) confirms the result noted in
Section~\ref{sec:nebvsstel}, namely that the ionized gas is on average
more obscured than the stellar continuum.  Specifically, we find that
the average difference in the magnitudes of attenuation between
recombination line and continuum photons at $6565$\,\AA\, is $\langle
A_{\rm H\alpha} - A_{\rm cont}\rangle \simeq 0.15$.  However, as was
the case in the comparison of the color excesses, we find that the
difference in total attenuation of line and continuum photons depends
on the SFR, such that galaxies with the largest SFRs can exhibit
line-emitting regions that are half a magnitude (or more) attenuated
than the stellar continuum.

\subsubsection{Physical Interpretation}

\begin{figure}[htb]
\epsscale{1.}
\plotone{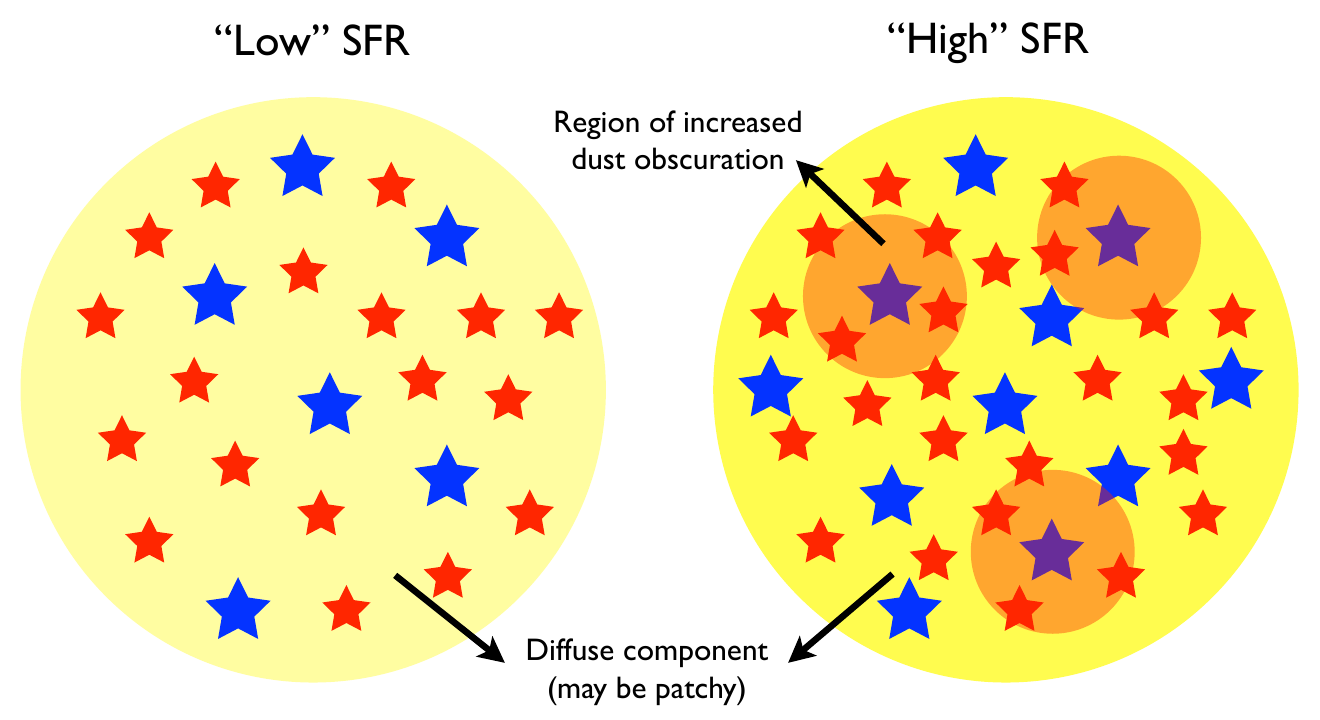}
\caption{Illustration of a simple geometry of dust and gas that can
  account for the trends in the difference between ionized gas and
  continuum color excess, and total attuation, versus SFR.  The yellow
  region denotes the diffuse dust component (that may be patchy).  The
  red regions indicate areas of increased dustiness within the galaxy.
  The blue and red stars indicate high mass (ionizing) and lower mass
  stars, respectively.  At lower SFRs ($\la
  20$\,$M_\odot$\,yr$^{-1}$), stars of all masses are uniformly
  obscured.  As the SFR increases, the diffuse component becomes more
  dust-enriched (as indicated by the darker shade of yellow), while
  regions of more highly obscured SFR (red regions) become prominent.
  As the SFR increases, these more obscured regions begin to dominate
  the nebular line and bolometric luminosities.  The diffuse component
  dominates the UV through optical SED at both low and high SFRs.}
\label{fig:cartoon_crop}
\end{figure}

\begin{figure}[htb]
\epsscale{1.}
\plotone{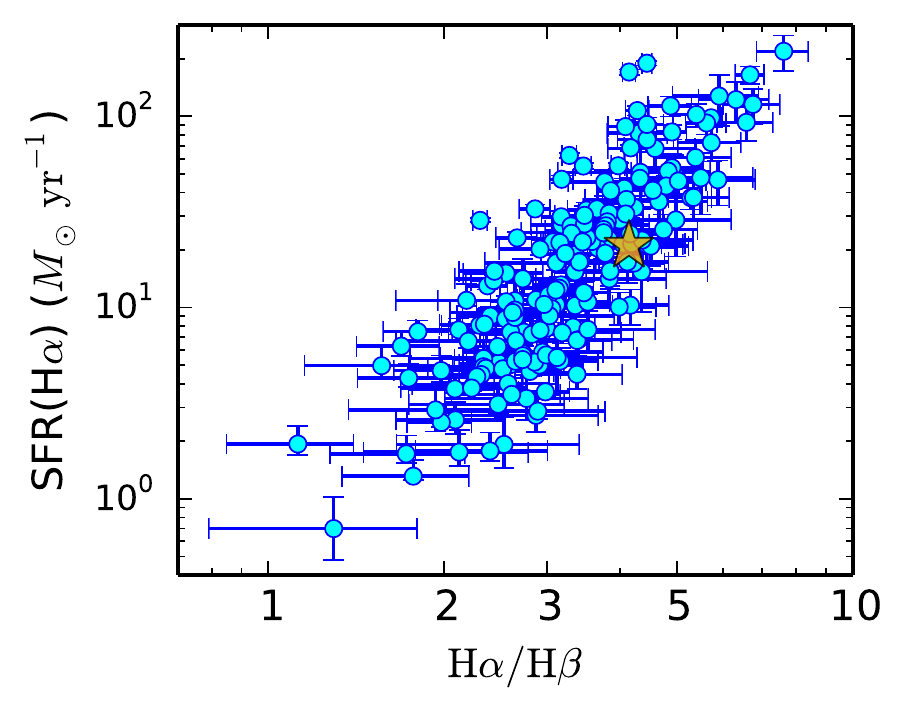}
\caption{H$\alpha$-based SFR versus Balmer decrement.  The large star
  denotes the average values for the H$\beta$-undetected galaxies.
  Note that the Balmer decrement is used in the calculation of the
  SFR, resulting in a tightening of the scatter shown here.
  Nonetheless, the results suggest the same trend between SFR and
  dustiness that has been noted in other studies at low and high
  redshift (see text).}
\label{fig:sfrvsdust}
\end{figure}

A final point of consideration is the strong dependence of total dust
attenuation on SFR (Figure~\ref{fig:sfrvsdust}).  Such a correlation
has been noted before both at low \citep{wang96, hopkins01} and high
redshift (e.g., \citealt{adelberger00, reddy06a, buat07, burgarella09,
  reddy08, buat09, reddy10,sobral12, dominguez13}), and has also been
shown to evolve with redshift (e.g., \citealt{reddy10, dominguez13}).
Physically, this trend can be understood as the product of increased
metal and dust enrichment of the ISM with increasing SFR.  In the
context of the present study, a {\em possible} simple model that can
account for the trends in ionized gas versus continuum attenuation is
one in which a modestly reddened stellar population dominates the
observed UV through optical continuum emission of galaxies over the
entire range of SFRs considered here.  Meanwhile, as the total SFR
rises, a greater fraction of this SFR is obscured in optically-thick
regions, as a consequence of the increased dust enrichment that
accompanies galaxies with larger SFRs.  As a result, the {\em average}
optical depth towards ionizing stars rises with SFR.  This simple
picture is illustrated in Figure~\ref{fig:cartoon_crop}.  We emphasize
that this is an ``average'' effect, as it is clear that some fraction
of O stars are in relatively unobscured regions of the galaxies where
they can contribute significantly to the UV continuum and where the
average reddening (as measured by the UV slope, $\beta$) is lower than
that measured from the Balmer lines.

The framework just discussed is somewhat different than the one
originally invoked to explain the disparity between the gas and
continuum color excesses in the local universe.  Locally, such a
discrepancy between color excesses is thought to arise from the
increased dust covering fraction surrounding HII regions (e.g.,
perhaps due to undissipated parent birth clouds surrounding the
youngest stars; \citealt{calzetti94}).  If this is the case at high
redshift, i.e., that {\em most} of the short-lived O stars reside in
or close to their parent molecular clouds, then we should see a {\em
  systematic} offset between gas and continuum color excesses at all
SFRs.  However, this expected systematic offset is not seen in our
data (e.g., Figures~\ref{fig:ebmvcompare}, \ref{fig:ebmvcompare_sfr}
and \ref{fig:compatt}).  Rather, our results suggest two distinct
stellar populations in each galaxy: one in which stars of all masses
are subject to modest amounts of dust reddening and which dominate the
observed UV through optical continuum luminosity; and another in which
stars of all masses reside in very dusty regions.  The second
population becomes more dominant over the first population (in terms
of its contribution to the nebular line and bolometric luminosities)
as the total SFR of a galaxy increases.  Consequently, the color
excesses (and total attenuation) of the gas and continuum diverge with
rising SFR.  This simple picture may also imply that the attenuation
curve relevant for the ionized gas may transition from a gray curve
(e.g., \citealt{calzetti00}) to a progressively ``steeper'' (e.g.,
Galactic or SMC) curve with increasing SFR.  

Along these lines, more recent high spatial resolution observations of
several nearby galaxies suggest that the relationship between the
attenuation of the gas and stars varies with SFR surface density
($\sum_{\rm SFR}$), such that the Balmer lines are systematically more
attenuated than the stellar continuum for regions where $\sum_{\rm
  SFR} \ga 0.01$\,$M_\odot$\,yr$^{-1}$\,kpc$^{-2}$ (e.g., see Figure~4
of \citealt{kreckel13}; see also \citealt{boquien15}).  These
observations have been interpreted as evidence that the Balmer lines
are dominated by dust-buried HII regions at high $\sum_{\rm SFR}$
\citep{kreckel13}.  If the higher SFR galaxies in our sample also have
higher $\sum_{\rm SFR}$, then the situation inferred locally on small
scales (e.g., 100-200\,pc) may also apply at high redshift.  This
issue can be addressed with the larger MOSDEF sample, which will
include galaxy size information.

Our analysis demonstrates that while the wavelength dependence of the
dust obscuration appears to be independent of specific SFR, the {\em
  total} obscuration of the ionized gas relative to that of the
stellar continuum varies with SFR and sSFR.  Specifically, galaxies
with larger SFRs (and sSFRs) exhibit redder color excesses and larger
total attenuations for the line emission relative to those for the
continuum.  These trends are responsible for the scatter seen between
$\beta$ and $\tau_{\rm b}$ (Figure~\ref{fig:beta_vs_taub}).  The
physical picture discussed here implies that galaxies with larger SFRs
may exhibit a higher dispersion in their spatially resolved colors as
the dust obscuration becomes more patchy with the rise of a second
(dustier) population.  Alternatively, the increased dust covering
fraction expected for galaxies with larger SFRs may be reflective of a
compact configuration of the star formation.  Hence, future prospects
for elaborating on this simple picture will include a detailed
analysis of how the SFR surface density, and the spatially resolved UV
colors of galaxies---whose measurement is aided with deep
high-resolution {\em HST} imaging in the fields targeted with
MOSDEF---correlate with globally-measured color excesses of the
ionized gas and the stellar continuum.

\subsection{Implications for SFR Measurements}
\label{sec:sfrimplications}

\subsubsection{H$\alpha$ and UV-based SFRs}

\begin{figure}
\epsscale{1.}
\plotone{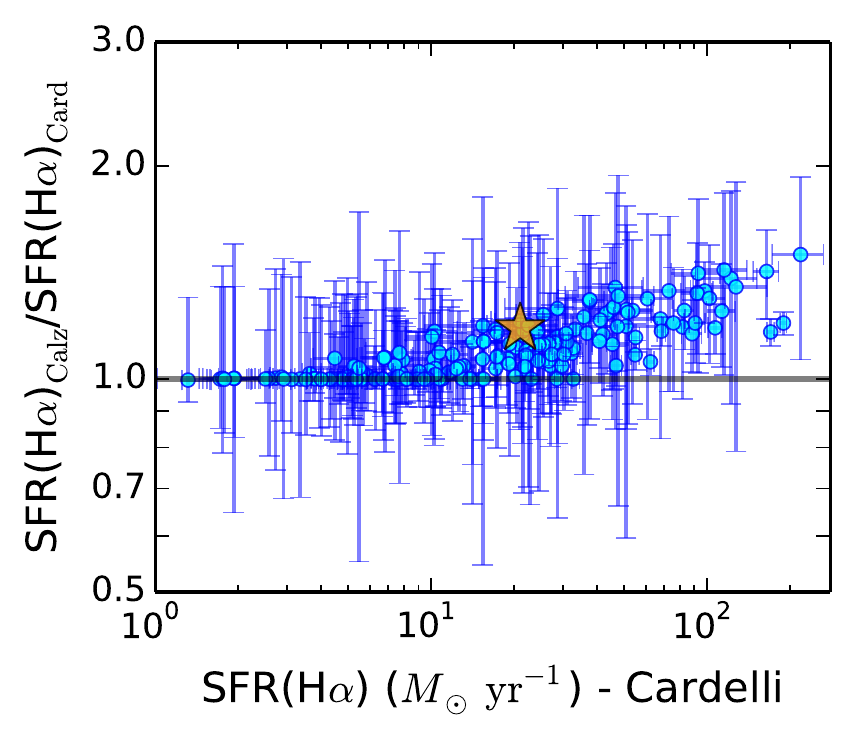}
\caption{Ratio of H$\alpha$-based SFRs assuming the \citet{calzetti00}
  attenuation curve and the \citet{cardelli89} extinction curve,
  versus the those assuming the \citet{cardelli89} extinction curve.
  The large star denotes the average values for the
  H$\beta$-undetected galaxies.  Assuming that the \citet{calzetti00}
  attenuation curve applies to the HII regions of high-redshift
  galaxies results in dust-corrected SFR(H$\alpha$) that can be as
  much as $\approx 50\%$ larger than those derived assuming the
  \citet{cardelli89} extinction curve.}
\label{fig:sfrhacompare}
\end{figure}

As mentioned in Section~\ref{sec:sedcompare}, the SED-based SFRs
assuming a MOSDEF attenuation curve are on average lower (by $\approx
20\%$) than those determined with the \citet{calzetti00} curve.  The
H$\alpha$-based SFRs also rely on an assumption of the attenuation
curve, both in deriving the color excess of the ionized gas, and in
applying the total dust correction to recover the intrinsic H$\alpha$
SFRs.  The default values of SFR(H$\alpha$) were computed assuming the
\citet{cardelli89} extinction curve ---i.e., this curve was used to
calculate $E(B-V)_{\rm gas}$ given the Balmer decrement, and the value
of the curve at $6565$\,\AA\, was used to dust correct the observed
H$\alpha$ luminosity (Section~\ref{sec:sfrcalc}).  The more common
practice of assuming that the \citet{calzetti00} curve applies to
high-redshift HII regions results in SFRs that can be as much as
$\approx 50\%$ larger than those assuming the \citet{cardelli89}
extinction curve (Figure~\ref{fig:sfrhacompare}), with an average
offset of $\approx 20\%$.  This comparison underscores the importance
of considering the appropriate attenuation curve to use when dust
correcting the line and continuum emission in high-redshift galaxies.

\begin{figure*}
\epsscale{1.}
\plotone{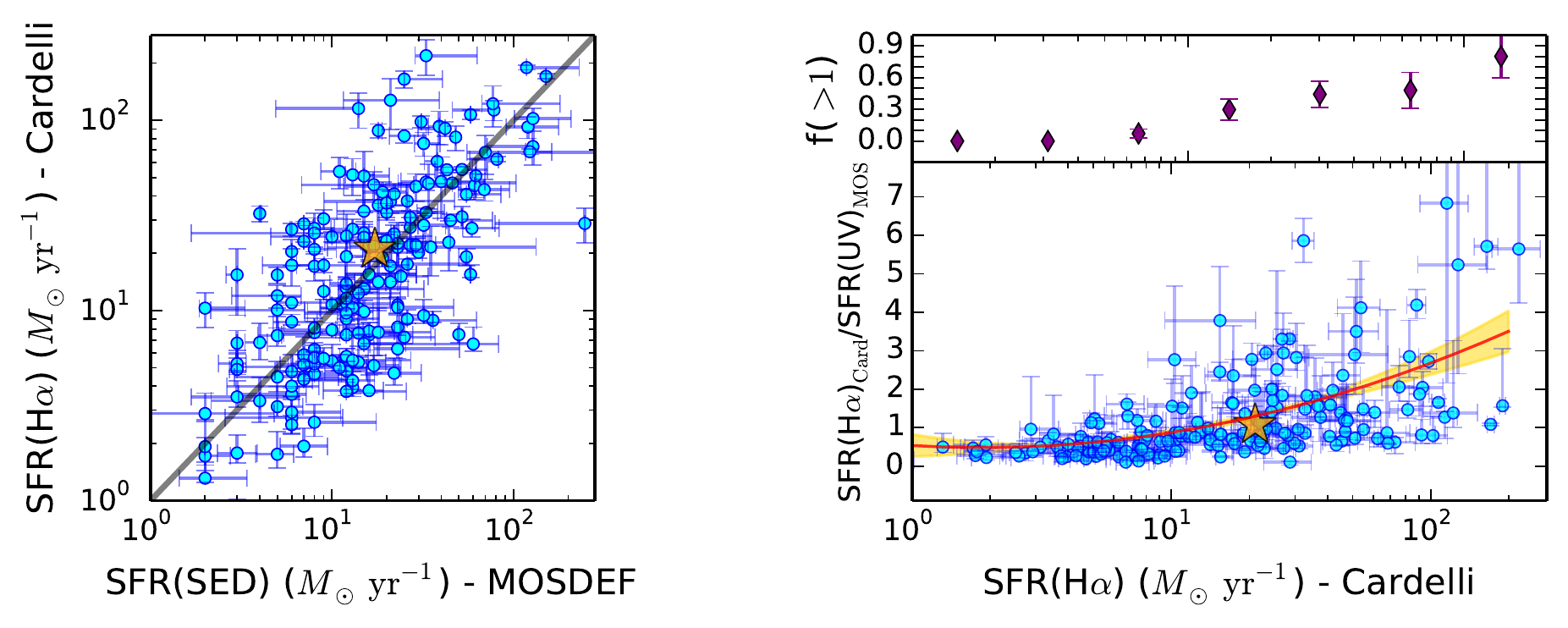}
\caption{{\em Left:} Comparison of SFR(H$\alpha$) assuming the
  \citet{cardelli89} extinction curve (and the Balmer decrement), and
  SED-based SFRs assuming the MOSDEF attenuation curve.  {\em Right
    (bottom):} Ratio of the dust-corrected H$\alpha$ and UV-based
  SFRs, plotted against SFR(H$\alpha$).  The UV-based SFR is
  calculated from the flux of the best-fit SED model at $1600$\,\AA,
  and corrected for dust assuming $E(B-V)_{\rm stars}$.  The red line
  and shaded region indicate the best-fit third-order polynomial fit
  and $68\%$ confidence interval between SFR(H$\alpha$)/SFR(UV) and
  $\log({\rm SFR(H\alpha)}/M_\odot\,{\rm yr}^{-1})$.  The large stars
  in both panels denote the average values for the H$\beta$-undetected
  galaxies.  {\em Right (top):} Fraction of galaxies where
  SFR(H$\alpha$) exceeds the UV-based SFR by more than 1\,$\sigma$, in
  bins of SFR(H$\alpha$).}
\label{fig:hasedcompare}
\end{figure*} 

When compared against each other, the H$\alpha$ and SED-determined
SFRs show a general agreement, with a scatter of $0.34$\,dex
(Figure~\ref{fig:hasedcompare}).  A similar agreement has been noted
in \citet{steidel14}, who also compare H$\alpha$ and SED-based SFRs,
but who use a different prescription for estimating the color excess
relevant for the nebular emission as they did not have Balmer
decrement measurements for their sample.  Notwithstanding this {\em
  general} agreement, the fraction of objects for which the
H$\alpha$-based SFR exceeds the UV- (or SED-) based SFR by more than
$1$\,$\sigma$ increases from $f=0$ to $f=0.8$ from
SFR(H$\alpha$)$\approx 1$\,$M_\odot$\,yr$^{-1}$ to
$300$\,$M_\odot$\,yr$^{-1}$.  We formalized this trend by fitting a
third-order polynomial to SFR(H$\alpha$)/SFR(UV) versus $x\equiv
\log[{\rm SFR(H\alpha)}/M_\odot\,{\rm yr}^{-1}]$:
\begin{equation}
\frac{\rm SFR(H\alpha)}{\rm SFR(UV)} = 0.29 + 0.31x + 0.29x^2 + 0.0040x^3.
\end{equation}
Figure~\ref{fig:hasedcompare} shows that $\ga 40\%$ of galaxies with
SFR(H$\alpha$)$\ga 20$\,$M_\odot$\,yr$^{-1}$ have UV- and SED-based
SFRs that underpredict the H$\alpha$-based SFR by an average factor of
$\simeq 2$.  It is unlikely that this trend is due to selection
effects, as the MOSDEF galaxy selection is based primarily on
rest-optical continuum emission, and the success of identifying a
redshift depends only on whether any of the rest-optical emission
lines are detected, independent of the UV continuum.  Thus, it is
improbable that we are missing a large number of galaxies where the
UV-based SFR is substantially smaller than the H$\alpha$-based SFRs
for SFR(H$\alpha$)$\la 20$\,$M_\odot$\,yr$^{-1}$.

\begin{figure}
\epsscale{1.}
\plotone{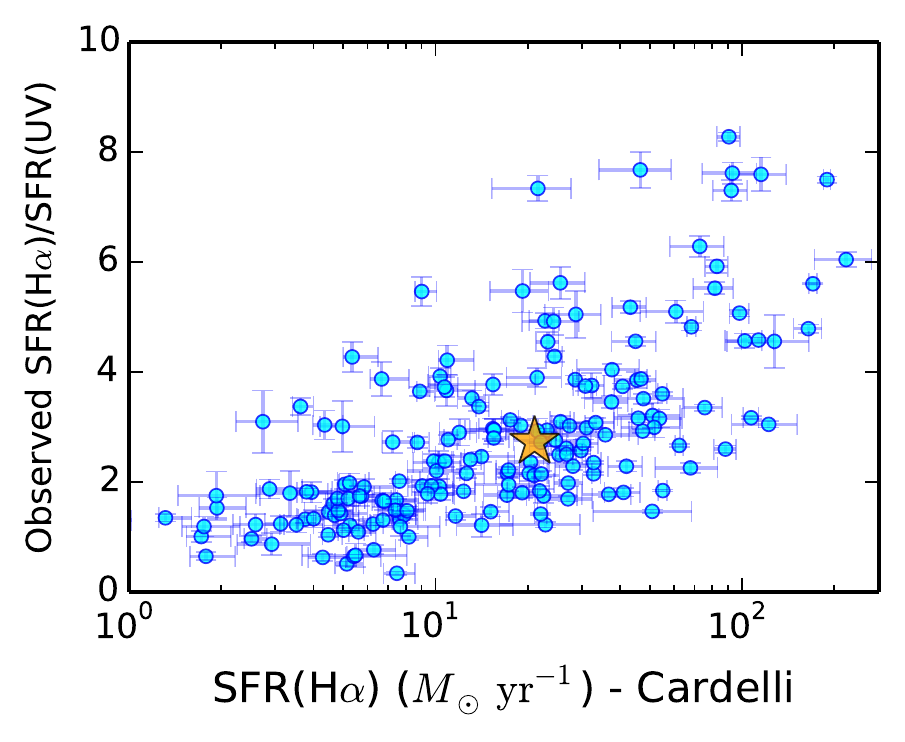}
\caption{Ratio of the observed H$\alpha$ and UV SFRs (i.e.,
  uncorrected for dust attenuation) versus the dust-corrected
  H$\alpha$ SFR.  The large star denotes the average values for the
  H$\beta$-undetected galaxies.  Those galaxies with dust-corrected
  SFR(H$\alpha$)$\la 10$\,$M_\odot$\,yr$^{-1}$ generally have observed
  SFR ratios that are consistent with unity within $3$\,$\sigma$.}
\label{fig:obssfr}
\end{figure} 

Focusing on those galaxies with lower SFRs, we find that their
SFR(H$\alpha$)/SFR(UV) ratios lie systematically below a value of
unity.  We must address whether this systematic difference is due to
variations in the dust attenuation of the nebular regions and stellar
continuum, or if there is some other mechanism that may be partly
responsible for the low SFR(H$\alpha$)/SFR(UV) ratios for these
galaxies.  To investigate this, we calculated the ratio of the {\em
  observed} H$\alpha$ and UV SFRs (i.e., uncorrected for dust
extinction) for all the objects in our sample
(Figure~\ref{fig:obssfr}).  The observed SFRs will include the light
emitted by unobscured stars, as well as the unattenuated light from
obscured stars.  Galaxies with (dust corrected) SFR(H$\alpha$)$\la
20$\,$M_\odot$\,yr$^{-1}$ have observed SFR(H$\alpha$)/SFR(UV) ratios
that are generally within $3$\,$\sigma$ of a ratio of unity.  As a
consequence, for the stars contributing to the observed H$\alpha$ and
UV luminosities, there is no evidence that we are observing these them
several tens of Myr after an initial burst of star formation: this
scenario would result in observed SFR ratios that deviate from unity
due to the different timescales (and mass ranges) over which the
H$\alpha$ and UV luminosities are sensitive to the total SFR.  Thus,
if the stars dominating the unobscured H$\alpha$ and UV luminosities
have a similar IMF and star-formation history as those stars that
dominate the obscured H$\alpha$ and UV luminosities, then the results
of Figure~\ref{fig:obssfr} suggest that the low dust-corrected
SFR(H$\alpha$)/SFR(UV) ratios measured for galaxies with
SFR(H$\alpha$)$\la 10$\,$M_\odot$\,yr$^{-1}$ are likely tied to
differences in the dust obscuration of the nebular emission and
stellar continuum, rather than differences in star-formation history,
changes in the IMF, or sampling stochasticity (c.f., \citealt{lee09,
  meurer09, fumagalli11, dasilva14, dominguez14}).

\subsubsection{A Prescription for Estimating $E(B-V)_{\rm gas}$ and
$A_{\rm H\alpha}$}
\label{sec:prescription}

\begin{figure}[htb]
\epsscale{1.}
\plotone{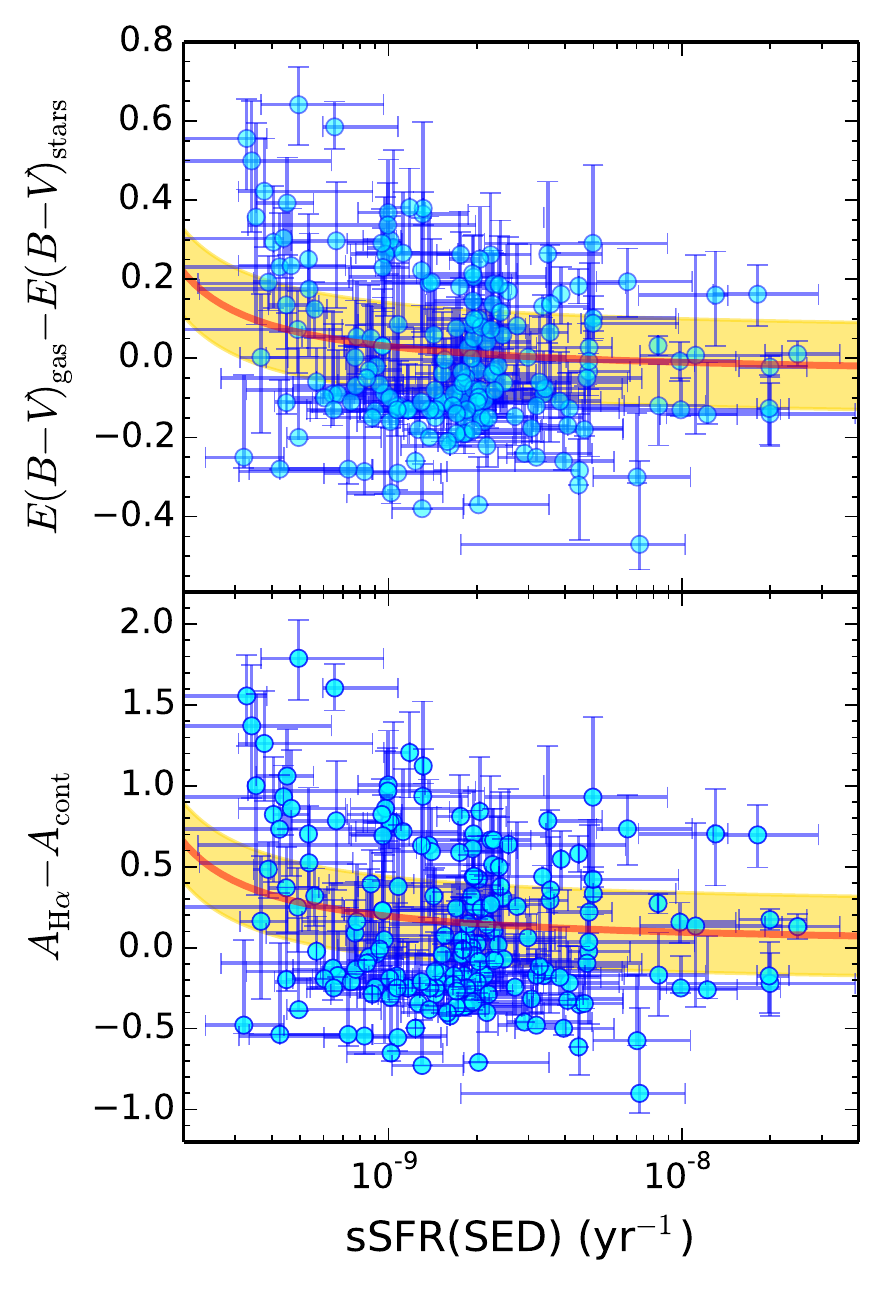}
\caption{Difference between the gas and continuum color excesses ({\em
    top}), and difference between the attenuation (in magnitudes) of
  recombination line and continuum photons ({\em bottom}) versus
  sSFR(SED).  Here, sSFR(SED) is computed using the SED-based SFR and
  $M^{\ast}$.  The solid lines and shaded regions denote the best-fit
  relations and the $1$\,$\sigma$ scatter, respectively, as indicated
  in the text.}
\label{fig:prescription}
\end{figure}

With our independent measurements of the color excesses of the stellar
continuum and ionized gas, we can quantify the relationship between
SED-inferred quantities (i.e., $E(B-V)_{\rm stars}$ and SFR(SED))
which are commonly available for high-redshift galaxy samples, and
$E(B-V)_{\rm gas}$, which can be difficult to constrain in the absence
of Balmer decrement measurements.  Figure~\ref{fig:prescription} shows
the difference between gas and continuum color excess ($E(B-V)_{\rm
  gas}-E(B-V)_{\rm stars}$), and the difference between the
attenuation of recombination and continuum photons ($A_{\rm H\alpha} -
A_{\rm cont}$), as a function of the SED-determined sSFR,
sSFR(SED). Note that sSFR(SED) combines the SED-determined SFR with
the SED-determined stellar mass, so it is not the same as the default
sSFR assumed throughout the paper (i.e., the latter combines
SFR(H$\alpha$) with $M^{\ast}$).  We chose to quantify these
relationships in terms of sSFR, and not SFR, as the correlations of
$E(B-V)_{\rm gas}-E(B-V)_{\rm stars}$ and $A_{\rm H\alpha} - A_{\rm
  cont}$ with sSFR(SED) are stronger than those with SFR(SED).  To
establish useful empirical relations between these quantities, we fit
them with functions of $\xi \equiv 1./(\log[{\rm sSFR(SED)}/{\rm
    yr}^{-1}] + 10)$:
\begin{equation}
E(B-V)_{\rm gas}-E(B-V)_{\rm stars} = -0.049 + 0.079/\xi,
\label{eq:ebmvprescrip}
\end{equation}
and 
\begin{equation}
A_{\rm H\alpha}-A_{\rm cont} = -0.0035 + 0.1972/\xi.
\label{eq:attprescrip}
\end{equation}
The scatter in these relations are $\sigma[E(B-V)_{\rm
    gas}-E(B-V)_{\rm stars}] \approx 0.12$ and $\sigma[A_{\rm
    H\alpha}-A_{\rm cont}] \approx 0.24$.  Given $E(B-V)_{\rm stars}$
and sSFR(SED) obtained from standard SED-fitting, one can use
Equations~\ref{eq:ebmvprescrip} and \ref{eq:attprescrip} to then {\em
  roughly} estimate $E(B-V)_{\rm gas}$ and $A_{\rm H\alpha}$, keeping
in mind the large scatter in these quantities.  Finally, note that the
difference in color excesses is found to {\em decrease} on average
with increasing sSFR(SED) (e.g., see also \citealt{price14}), {\em
  opposite} of the trend that we found between $E(B-V)_{\rm
  gas}-E(B-V)_{\rm stars}$ and the default sSFRs that assume
H$\alpha$-based SFRs (Figures~\ref{fig:ebmvcompare} and
\ref{fig:compatt}).  This behavior simply reflects the systematic
difference between SED-based and H$\alpha$-based SFRs, as discussed
above.

\subsubsection{Summary of SFR Comparisons}

The physical picture put forth in the previous section leads to the
expectation that UV- (or SED-) based SFRs should diverge from
H$\alpha$-based ones---which we consider to be more reliable than the
former---for galaxies with large SFRs, as an increasing fraction of
the nebular line and bolometric luminosities arise from regions of
higher dust covering fraction.  The statistics indicating an
increasing SFR(H$\alpha$)/SFR(UV) ratio with SFR(H$\alpha$) are
consistent with this scheme.  A similar effect is seen in the
comparison of UV and IR-based SFRs (e.g., \citealt{goldader02,
  reddy10, howell10}), where the former generally underpredict the
latter for galaxies where most of the star formation occurs in
optically-thick regions that are essentially ``invisible'' in the UV.
Because the attenuation curve derived for the continuum is most
sensitive to the less-reddened stellar populations in a galaxy, it is
not surprising that inferences of total SFRs based on the UV continuum
emission will underestimate the total SFR for very dusty galaxies.
The assumption of UV- or SED-based SFRs may also lead to erroneous
conclusions regarding the dependence of the difference in color excess
between the stars and ionized gas, and the sSFR (e.g.,
Figures~\ref{fig:ebmvcompare} and \ref{fig:prescription}).

Thus, the common practice of assuming either equal color excesses of
the ionized gas and stellar continuum, or assuming that the ionized
gas color excess is systematically larger than that of the stellar
continuum, may lead to incorrect inferences of the SFRs.  We also
highlight the importance of treating the nebular emission and stellar
continuum with the appropriate attenuation/extinction curves.  Here,
we have made independent measurements of the color excess of the
ionized gas (aided by the Balmer decrement), and have calculated the
the attenuation curve relevant for the stellar continuum.  Based on
these analyses, we find disparities between the H$\alpha$- and
UV-based SFRs that signify a varying contribution of dusty stellar
populations to the bolometric luminosities of high-redshift galaxies.

\section{Conclusions}
\label{sec:conclusions}

We use a sample of 224 star-forming galaxies at redshifts
$z=1.36-2.59$ with measurement of H$\alpha$ and H$\beta$ emission
lines from the MOSDEF survey to investigate the effects of dust on the
stellar continuum of high-redshift galaxies.  Our analysis suggests
that the reddening of the UV continuum with increasing Balmer line
opacity is most directly tied to dust obscuration.  We calculate the
attenuation curve by constructing, and taking ratios of, the composite
photometric measurements of galaxies in bins of specific SFR
and Balmer optical depth.  The attenuation curve derived here is very
similar in shape to that of the SMC curve at $\lambda \ga 2500$\,\AA.
Blueward of this wavelength, the curve has a shape that is identical
to, but with a normalization that is $\Delta k(\lambda)\simeq 1.63$
lower than, that of the \citet{calzetti00} attenuation curve.
Moreover, we find a marginal trend between the depth of absorption at
$2175$\,\AA\, and the color excess of the stellar continuum.

We explore the implications for the attenuation curve on the stellar
populations and SFRs of high-redshift galaxies.  In general, the new
attenuation curve implies slightly redder color excesses ($E(B-V)$),
$\approx 20\%$ lower SFRs, and stellar masses that are
$\Delta\log(M^{\ast}/M_\odot) \approx 0.16$\,dex lower, than those
computed using the typically assumed starburst attenuation curve
\citep{calzetti00}.  We also find that while the color excess of the
ionized gas is similar to that of the stellar continuum for about half
of the objects in our sample, the difference in the reddening---and in
the total attenuation---of the gas and stars in high-redshift galaxies
is a strong function of SFR.  In particular, we find that the ionized
gas is more reddened relative to the stellar continuum with increasing
SFR. Our results are consistent with a simple picture in which the UV
through optical continuum of high-redshift galaxies is dominated by a
modestly reddened stellar population, while a second, dustier, stellar
population begins to dominate the bolometric luminosity and nebular
line luminosity with increasing SFR.  In this picture, UV- and
SED-based SFRs may diverge from (i.e., underpredict) the total SFR
even for galaxies with relatively modest H$\alpha$-based SFRs ($\ga
20$\,$M_\odot$\,yr$^{-1}$).

Our analysis and results demonstrate the utility of using high quality
near-IR spectroscopic data, along with multi-wavelength photometry, to
directly measure the attenuation curve at high redshift.  Further
spectroscopy from the ongoing MOSDEF survey, along with deep mid- and
far-IR imaging data, will allow us to build upon these results, and
will be crucial for removing a key uncertainty (namely the shape and
normalization of the attenuation curve) in the measurement of
star-formation rates and the interpretation of stellar populations in
the distant Universe.

\acknowledgements

NAR thanks Chuck Steidel, Lee Armus, Daniela Calzetti, and Max Pettini
for feedback on the manuscript.  We thank the referee for a careful
reading of the manuscript and suggestions for clarifying the text.  We
acknowledge support from NSF AAG grants AST-1312780, 1312547, 1312764,
and 1313171.  We are grateful to the MOSFIRE instrument team for
building this powerful instrument, and to Marc Kassis at the Keck
Observatory for his many valuable contributions to the execution of
the MOSDEF survey.  We also acknowledge the 3D-HST collaboration, who
provided us with spectroscopic and photometric catalogs used to select
MOSDEF targets and derive stellar population parameters.  We also
thank I. McLean, K. Kulas, and G. Mace for taking observations for the
MOSDEF survey in May and June 2013.  NAR is supported by an Alfred
P. Sloan Research Fellowship.  MK acknowledges support from a
Committee Faculty Research Grant and a Hellmann Fellowship.  ALC
acknowledges funding from NSF CAREER grant AST-1055081.  We wish to
extend special thanks to those of Hawaiian ancestry on whose sacred
mountain we are privileged to be guests.  Without their generous
hospitality, most of the observations presented herein would not have
been possible.



\begin{appendices}

\end{appendices}

\clearpage



\end{document}